\documentclass[prd,letterpaper,superscriptaddress,reprint,nofootinbib]{elsarticle} 

\usepackage{floatrow}
\usepackage{amssymb}
\usepackage{amsthm}
\usepackage{amsmath}
\usepackage{amsbsy}
\usepackage{nicefrac}
\usepackage{wrapfig}
\usepackage{lineno}
\usepackage[caption=false]{subfig}
\usepackage{natbib}
\usepackage{graphicx}
\include{rlFig}
\graphicspath{{./}{fig/}}

\begin{document}

\title{Antarctic Surface Reflectivity Calculations and Measurements from the ANITA-4 and HiCal-2 Experiments}

\author[KU]{S.~Prohira}
\author[KU,MEPHI]{A.~Novikov}
\author[IITT]{P.~Dasgupta}
\author[IITT]{P.~Jain}
\author[IITT]{S.~Nande}
\author[OSU,CCAP]{P.~Allison}
\author[OSU,CCAP]{O.~Banerjee}
\author[UCL]{L.~Batten}
\author[OSU,CCAP]{J.~J.~Beatty}
\author[JPL]{K.~Belov}
\author[KU,MEPHI]{D.~Z.~Besson}
\author[WashU]{W.~R.~Binns}
\author[WashU]{V.~Bugaev}
\author[UD]{P.~Cao}
\author[NTU]{C.~Chen}
\author[NTU]{P.~Chen}
\author[UD]{J.~M.~Clem}
\author[OSU,CCAP]{A.~Connolly}
\author[UCL]{L.~Cremonesi}
\author[OSU]{B.~Dailey}
\author[Chicago]{C.~Deaconu}
\author[UCLA]{P.~F.~Dowkontt}
\author[UH]{B.~D.~Fox}
\author[OSU]{J.~Gordon}
\author[UH]{P.~W.~Gorham}
\author[SLAC]{C.~Hast}
\author[UH]{B.~Hill}
\author[OSU]{R.~Hupe}
\author[WashU]{M.~H.~Israel}
\author[UCLA]{J.~Lam}
\author[NTU]{T.~C.~Liu}
\author[Chicago]{A.~Ludwig}
\author[UH]{S.~Matsuno}
\author[UH]{C.~Miki}
\author[UCL]{M.~Mottram}
\author[UD]{K.~Mulrey}
\author[NTU]{J.~Nam}
\author[UCL]{R.~J.~Nichol}
\author[Chicago]{E.~Oberla}
\author[KU]{K.~Ratzlaff}
\author[WashU]{B.~F.~Rauch}
\author[JPL]{A.~Romero-Wolf}
\author[UH]{B. Rotter}
\author[UH]{J.~Russell}
\author[UCLA]{D.~Saltzberg}
\author[UD]{D.~Seckel}
\author[UH]{H.~Schoorlemmer}
\author[OSU]{S.~Stafford}
\author[KU]{J.~Stockham}
\author[KU]{M.~Stockham}
\author[UCL]{B.~Strutt}
\author[UH]{K.~Tatem}
\author[UH]{G.~S.~Varner}
\author[Chicago]{A.~G.~Vieregg}
\author[CalPoly]{S.~A.~Wissel}
\author[UCLA]{F.~Wu}
\author[KU]{R.~Young}

\address[UCLA]{Dept. of Physics and Astronomy, Univ. of California, Los Angeles, Los Angeles, CA 90095.}
\address[OSU]{Dept. of Physics, Ohio State Univ., Columbus, OH 43210.}
\address[UH]{Dept. of Physics and Astronomy, Univ. of Hawaii, Manoa, HI 96822.}
\address[NTU]{Dept. of Physics, Grad. Inst. of Astrophys.,\& Leung Center for 
Cosmology and Particle Astrophysics, National Taiwan University, Taipei, Taiwan.}
\address[UCI]{Dept. of Physics, Univ. of California, Irvine, CA 92697.}
\address[KU]{Dept. of Physics and Astronomy, Univ. of Kansas, Lawrence, KS 66045.}
\address[WashU]{Dept. of Physics, Washington Univ. in St. Louis, MO 63130.}
\address[SLAC]{SLAC National Accelerator Laboratory, Menlo Park, CA, 94025.}
\address[UD]{Dept. of Physics, Univ. of Delaware, Newark, DE 19716.}
\address[UCL]{Dept. of Physics and Astronomy, University College London, London, United Kingdom.}
\address[UMinn]{School of Physics and Astronomy, Univ. of Minnesota, Minneapolis, MN 55455.}
\address[JPL]{Jet Propulsion Laboratory, Pasadena, CA 91109.}
\address[CCAP]{Center for Cosmology and Particle Astrophysics, Ohio State Univ., Columbus, OH 43210.}
\address[Chicago]{Dept. of Physics, Enrico Fermi Institute, Kavli Institute for Cosmological Physics, Univ. of Chicago , Chicago IL 60637.}
\address[GSFC]{Currently at NASA Goddard Space Flight Center, Greenbelt, MD, 20771.}
\address[CalPoly]{Dept. of Physics, California Polytechnic State Univ., San Luis Obispo, CA 93407.}
\address[IITT]{Dept. of Physics, Indian Institute of Technology, Kanpur, Uttar Pradesh 208016, India}
\address[MEPHI]{National Research Nuclear University, Moscow Engineering Physics Institute, 31 Kashirskoye Highway, Rossia 115409}

\date{\today}

\message{<azimuth> for D-paired vs D-unpaired events}

\message{Effect of crust on surface reflectivity esp. at high obliquity}
\message{Fresnel zone at high obliquity becomes an ellipse extending how many degrees in either direction - must calculate this!}

\begin{abstract}
The balloon-borne HiCal radio-frequency (RF) transmitter,
in concert with the ANITA radio-frequency receiver array,  
is designed to measure the Antarctic surface reflectivity 
in the RF wavelength regime. The amplitude of surface-reflected
transmissions from HiCal, registered as triggered events 
by ANITA, can be compared with the direct transmissions preceding them
by $\cal{O}$(10) microseconds, to infer the surface power reflection coefficient $\cal{R}$. The first HiCal mission 
(HiCal-1, Jan. 2015) yielded
a sample of 100 such pairs, resulting in estimates of $\cal{R}$ at highly-glancing angles (i.e., zenith angles approaching $90^\circ$), with
 measured reflectivity for those events which exceeded extant calculations\citep{gorham2017antarctic}.
The HiCal-2 experiment, flying from Dec., 2016--Jan., 2017, provided 
an improvement by nearly two orders of magnitude in our event statistics, allowing a considerably
more precise mapping of the reflectivity over a wider range of incidence angles.
We find general agreement between the HiCal-2 reflectivity results and those obtained with the
earlier HiCal-1 mission, as well as estimates from 
Solar reflections in the radio-frequency regime\citep{besson2015antarctic}.
In parallel, our calculations of expected reflectivity have matured; herein, we use a plane-wave
expansion to estimate the reflectivity $\cal{R}$
from both a flat, smooth surface (and, in so doing, recover the
Fresnel reflectivity equations) and also a curved surface.
Multiplying our flat-smooth reflectivity by improved 
Earth curvature and surface roughness corrections now provides significantly better agreement between
theory and the HiCal 2a/2b measurements.
\end{abstract}
\maketitle

\section{Overview}
The NASA-sponsored ANITA project\citep{gorham2010observational,GorhamAllisonBarwick2009,Gorham:2016zah,Gorham:2010kv} 
has the goal of detecting the highest-energy particles incident on the Earth.
Although designed for measurement of ultra-high energy neutrinos interacting in-ice, the first ANITA flight
also demonstrated (unexpectedly) excellent sensitivity to primary
ultra-high energy cosmic rays (UHECR) with energies exceeding 1 EeV ($10^{18}$ eV)\citep{Hoover:2010qt}
interacting in the
Earth's atmosphere. These are assumed to be charged nuclei (likely protons), given the lack of efficient
acceleration mechanisms for electrically uncharged particles, and the long lifetimes required to 
traverse megaparsec-scale distances. Through interactions with terrestrial matter,
both neutrinos and charged cosmic-rays 
produce observable radio-frequency (RF) emissions via the Askaryan Effect\citep{Askaryan1962a,Askaryan1962b,Askaryan1965}, with three important distinctions between the two experimental
signatures:
\begin{enumerate}
\item as viewed from the airborne ANITA gondola, 
charged primary cosmic ray interactions in the atmosphere generally produce down-coming signals, which subsequently reflect
off the surface and up to the gondola, whereas
neutrinos interacting in-ice produce up-coming signals which refract through the surface to ANITA.
\item owing to the relative sparseness of the air target medium, 
down-coming charged cosmic ray interactions
 result in forward-beamed RF signal close to the primary CR momentum axis (within one degree), whereas Cherenkov radiation from
in-ice neutrino interactions is well-separated ($\theta_C\sim 57^\circ$) from the neutrino momentum axis.
\item down-coming charged primary cosmic-rays, owing to the ${\vec v}\times{\vec B}$ Lorentz force in the
Earth's magnetic field, result in predominantly horizontally-polarized (HPol) radiation, whereas the 
measurable Cherenkov radiation due to neutrino interactions and emerging to the gondola 
is predominantly vertically-polarized (VPol).
Owing to this latter consideration, the HPol component of the ANITA trigger was, unfortunately, removed
from the trigger chain after the ANITA-1 flight, and before the ANITA-2 flight (and also before it was realized that
ANITA-1 had charged UHECR measurement capabilities). That capability was re-installed for ANITA-3 and 
subsequent flights.
\end{enumerate}

In both cases, knowledge of the RF reflection/transmission across the 
surface discontinuity between Antarctic snow and air is critical to reconstructing
UHECR energies. This quantity is primarily determined by the dielectric contrast across the
discontinuity and also surface roughness effects, which can introduce, as a function of
signal incidence angle,
frequency-dependent decoherence and/or frequency-dependent signal amplification. At highly oblique
incidence angles, the divergence of signal upon reflection from the convex Earth surface results in a
significant dimunition of measured signal (i.e., ``curvature effects'').
Previously, the surface reflectivity was deduced from both ANITA-2\citep{besson2015antarctic} and also ANITA-3
observations of the Sun, and also ANITA-3 measurements of HiCal-1 triggers\citep{gorham2017antarctic}. 
Those measurements
typically followed expectations from the Fresnel equations, with the exception of the
most oblique incidence angles, for which HiCal-1 data indicated a two-fold larger-than-expected
surface reflectivity, compared to published models\citep{Schoorlemmer:2015afa}. 

Our goals for the successor HiCal-2 experiment, compared to HiCal-1 were three-fold: a) improvement of 
event statistics by at least an order of magnitude, b) considerably
greater incidence angle sampling
than the limited range probed by HiCal-1 (3.5--5 degrees with respect to the surface), 
and extension into the 8--30 degree incidence
angle regime probed by the Solar surface reflectivity measurements, and c) signal-emission
time-stamping and azimuthal orientation readback. The latter
is important in understanding the signal strength received at ANITA, given the expected dipole
beam pattern of the transmitter.

In what follows, we first detail the hardware used on the
HiCal-2 payloads (designated ``a'' and ``b'', in reverse order of
launch), as well as provide flight trajectory performance characteristics.
More details on the instrument can be found elsewhere\citep{gorham2017hical}.
We also provide details on our improved 
calculation of the expected surface reflectivity, and compare with our 
measured reflectivity.

\section{HiCal payload}
The HiCal payload schematically consists of three main components. These are:
\begin{enumerate}
\item  the Micro-Instrumentation Package (MIP)
containing the Columbia Scientific Ballooning Facility (CSBF) 
hardware used for communications with the main ground station in 
Palestine, TX, and also instrumentation for monitoring in-flight
payload and telemetry of useful data,
\item a sealed pressure vessel (PV) containing the bicone transmitter antenna,
two piezo-electric signal generators at each axial end of the bicone 
transmitter, and, for each piezo, both a rotating camshaft which activates the piezo every 8--10 seconds
(depending on voltage applied to the rotor, as well as ambient temperature) and also wires connecting the piezo to the
feedpoint at the center of the HiCal biconical antenna. Note that the camshaft period
for HiCal-1 was considerably shorter than for HiCal-2, of order 2.5--3 seconds.
The PV maintains a roughly
$\sim$1000 mb environment (compared to $\sim$5 mb outside the payload at float altitude).
Owing to the increased likelihood of high-voltage breakdown with decreasing pressure, the PV 
is essential in ensuring regular pulsing and reproducible signal shapes.
\item the Azimuth, TimeStamp and Altitude (ATSA) board which provides 
information on the transmitter performance in-flight, and measures the time and azimuthal 
transmitter orientation at the 
time a HiCal pulse is emitted.
\end{enumerate}
Given the 60,000 $ft^3$ balloon used to fly the payload, lift is sufficient to 
accommodate, at most, a total mass of 5 kg, similar to the weight limit on a typical weather balloon flight.

\subsection{Telemetry}
Since the HiCal payloads are not recovered, all necessary data are only retrievable via real-time
transmission at flight-time.
HiCal data are telemetered via the Iridium-based satellite communications network, with a total telemetry
bandwidth budget limited to 256 bytes/minute\footnote{The 256 byte/minute total science telemetry therefore restricted
the duty cycle for
transmission of ATSA information to less than 50\%, with the remainder being taken up by CSBF MIP data.}.
In addition to the time and azimuthal information from the ATSA 
board, CSBF data also include the voltage on the MIP board itself, as well as voltages monitoring the PV pressure
and the voltage being delivered to the piezo cam motor. 

\subsection{ATSA performance}
\begin{figure}[!ht]\begin{floatrow}
\ffigbox[\FBwidth]{\includegraphics[width=0.5\textwidth]{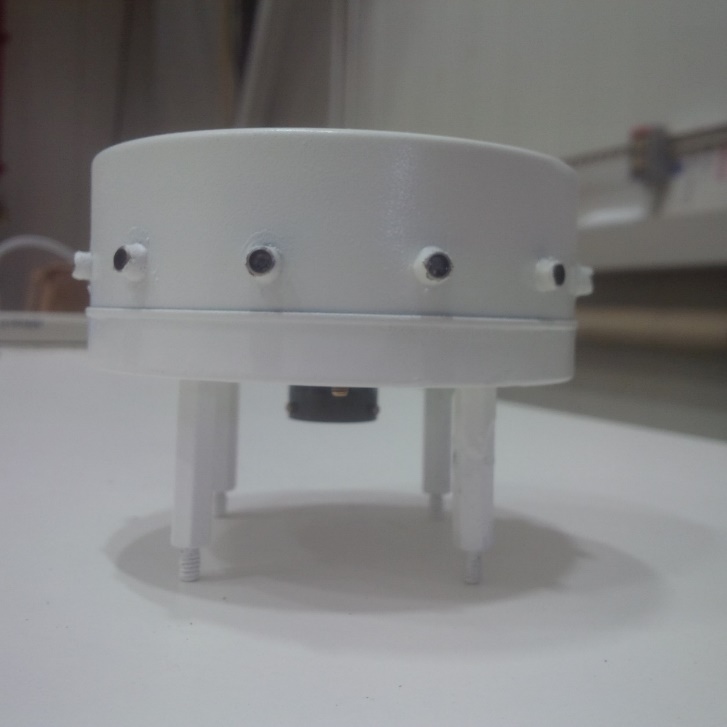}}{\caption{\it Photograph of ATSA sun sensor, employing 12 MEPhI-mark Silicon Photomultipliers (SiPM) arranged azimuthally.}\label{fig:atsa-photo}}
\ffigbox[\FBwidth]{\includegraphics[width=0.5\textwidth]{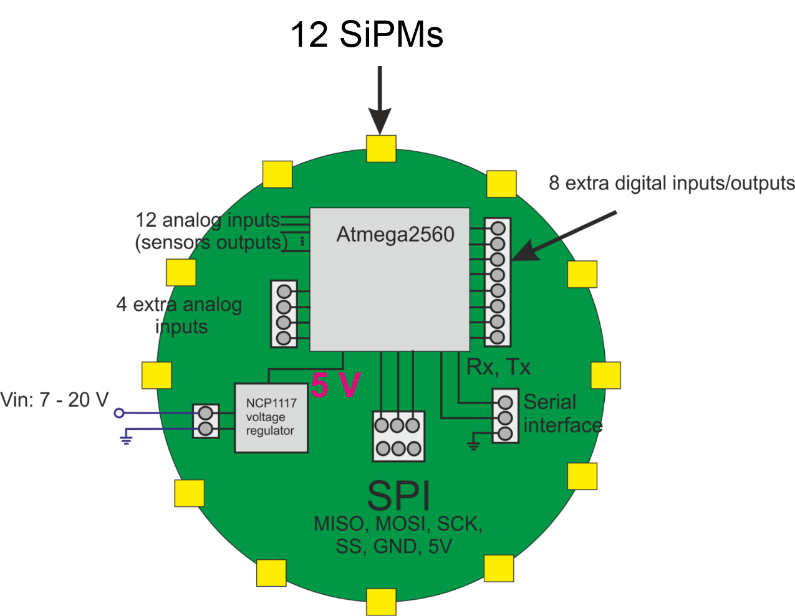}}{\caption{\it ATSA functional block diagram.}\label{fig:atsa_schematic}}
\end{floatrow}\end{figure}

\begin{figure}[!ht]\begin{floatrow}
\ffigbox[\FBwidth]{\includegraphics[width=0.5\textwidth]{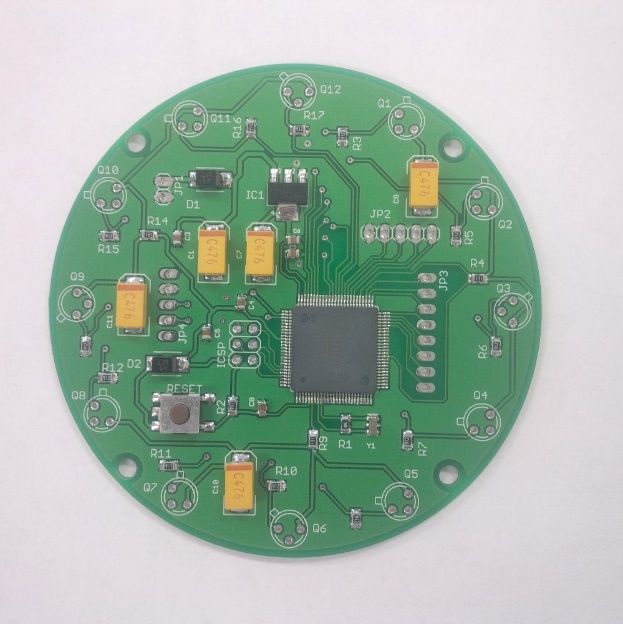}}{\caption{\it ATSA printed circuit board}\label{fig:atsa-pic}}
\ffigbox[\FBwidth]{\includegraphics[width=0.5\textwidth]{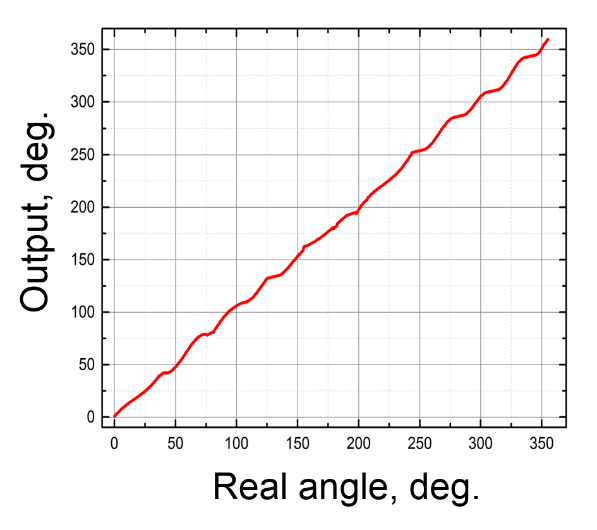}}{\caption{\it Calibration of azimuthal response of ATSA board}\label{fig:atsa-az}}
\end{floatrow}\end{figure}

To determine the absolute azimuthal orientation of the HiCal-2 payload,
the ATSA board interpolates the amplitudes of Solar-induced signals 
measured in 12 silicon photomultipliers (SiPM's) prepared prior to flight by
collaborators at the Moscow Engineering Physics Institute (Moscow, Russia). 
Each SiPM is displaced by 30 degrees, on a disk, relative to the
adjacent SiPM's, as shown in Figure \ref{fig:atsa-photo}. The
azimuthal orientation of the transmitter antenna axis is then calculated using
an ephemeris look-up table of the true sky location of the Sun, given the instantaneous
payload UTC time, latitude, longitude and elevation.
The block functional diagram and the actual implementation of the block diagram are shown in
Figures \ref{fig:atsa_schematic} and 
\ref{fig:atsa-pic}, respectively. 

Calibration of the Solar azimuthal response and the corresponding angular resolution 
is conducted on a bright Midwestern day. 
As seen from the deviation from linearity with unit slope
(Figure \ref{fig:atsa-az}), the ATSA azimuthal calibration has an accuracy of approximately three degrees. Note that there
is no tracking of the polar attitude of the payload, although measurements of the (albeit much heavier) ANITA gondola rarely show departures from horizontal exceeding one degree.

GPS time is provided by the CSBF MIP board; when a HiCal signal is produced by relaxation of the 
cam-depressed piezo, a small wire
pickup within the pressure vessel forwards this signal to the ATSA board, which then 
latches the CSBF GPS second and interpolates the
sub-second by counting clock cycles on a 200 MHz oscillator. 
This procedure was tested pre-flight and indicated
30 microsecond resolution. The actual in-flight timing resolutions achieved by
the ATSA were found to be approximately 30 $\mu$s and 500 $\mu$s for HiCal-2a and HiCal-2b, respectively, by
comparing HiCal time stamps to those of ANITA for the HiCal events which triggered ANITA-4.
The error in the timing for HiCal-2a is found to be dominated by jitter in the capacitive pickup latching the
CSBF GPS board. The source of the large timing error for HiCal-2b is due, in part, to the less distinctive HiCal-2b
observed output signal.

\subsection{Piezo-based transmitter}
For the HiCal-2 mission, three transmitter design modifications were employed relative to HiCal-1. 
First, the MSR-brand 
piezo-electric was selected to replace the previous HiCal-1b MHP piezo-electric, based on a lab
study of signal
shape and signal regularity. In pre-flight laboratory testing, the MSR brand piezo consistently produced
5 Volt (peak-peak) amplitude signals, when broadcast to an ANITA-2 Seavey quad-ridged horn antenna at a distance of 20 meters, 
with no additional amplification, translating to $\sim$5 kV signal output at the bicone transmitter antenna itself. 
Second (as mentioned earlier), to provide redundancy, each antenna was equipped with two piezo-electric
generators, one at either end of the dipole antenna. 
Finally, to minimize weight, the RICE dipole transmitter which flew in HiCal-1b was replaced
with the thinner, aluminium bicone model, with the separation between the two bicone halves reduced to
250 microns using a thin nylon spacer. 

As with all CSBF missions, prior to Antarctic flight, the performance of experimental hardware was
verified during the pre-flight summer in Palestine, TX. A photograph of the payload pressure vessel (black cylinder)
suspended beneath the MIP box is shown in Fig. \ref{fig:hangtestfoto}.

\section{Flight Details and Trajectory}
\begin{figure}[htpb]\begin{floatrow}
\ffigbox[\FBwidth]{\includegraphics[scale=0.8]{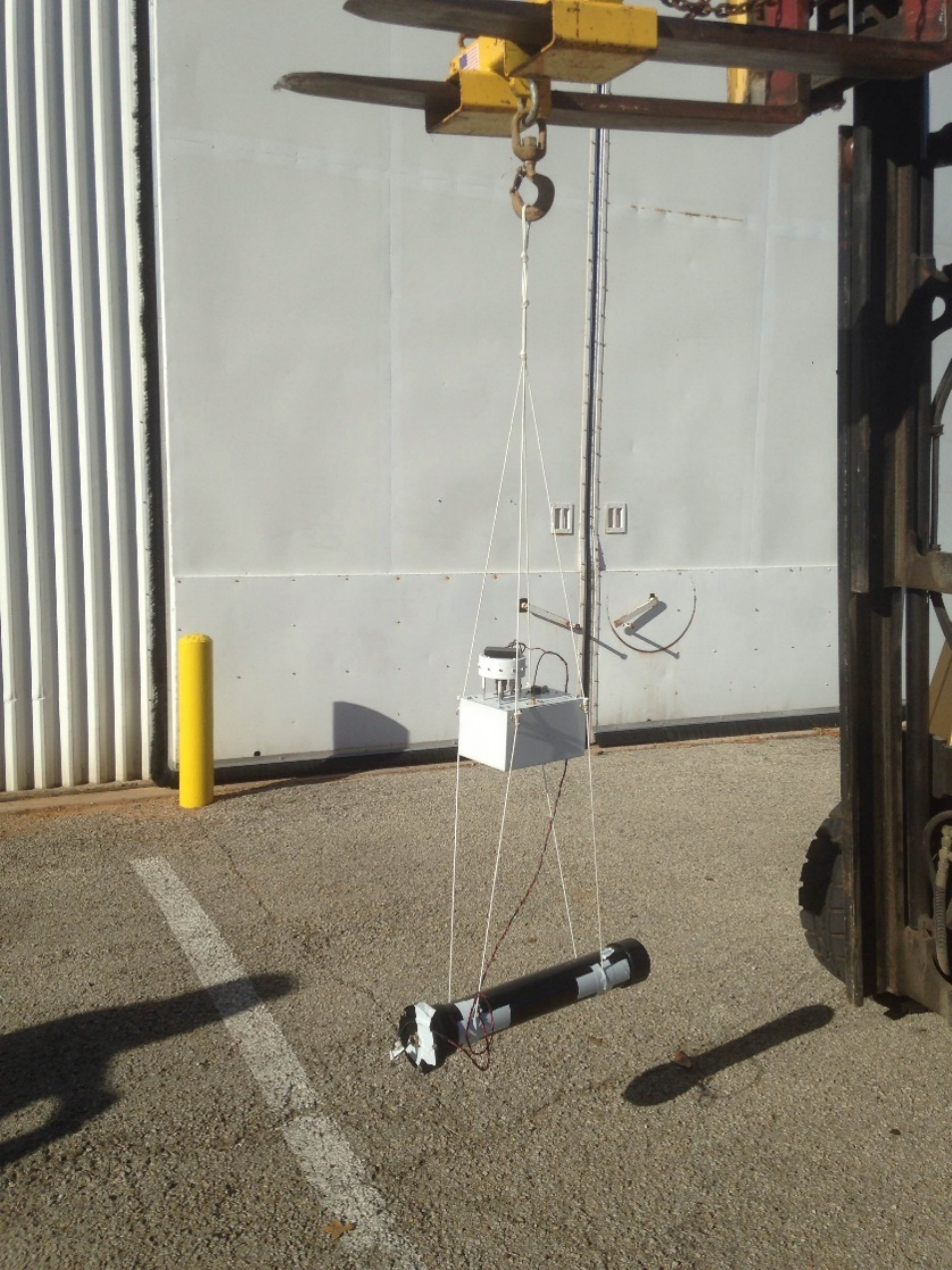}}{\caption{\it Photo of HiCal payload taken during pre-flight hang test (June, 2016, Palestine, TX).}\label{fig:hangtestfoto}}
\ffigbox[\FBwidth]{\includegraphics[width=0.5\textwidth]{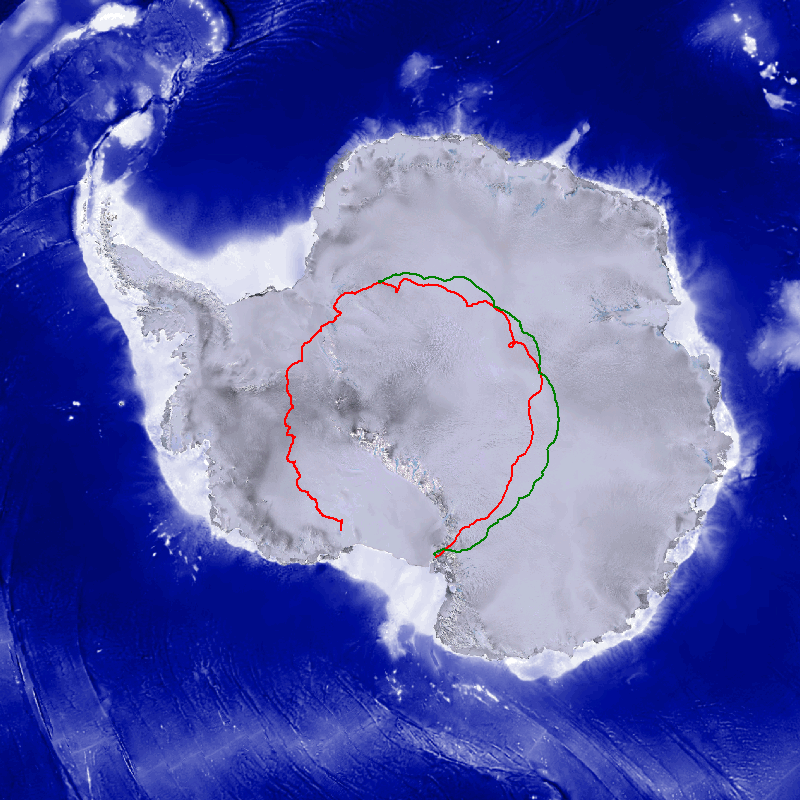}}{\caption{\it HiCal-2b payload (red) vs. HiCal-2a payload (green) trajectories. Note that these trajectories correspond to only those times when there was sufficient battery voltage within the MIP to telemeter GPS timestamps and also power the HiCal transmitter.}\label{fig:balloon1both}}
\end{floatrow}\end{figure}

Although originally intended to launch directly following the ANITA-4 launch on Dec. 2, 2016, logistical
restrictions made this impossible, and the decision 
was made to delay HiCal launch until the return of ANITA-4 to McMurdo Station following one full circumpolar orbit around
the Antarctic continent. HiCal-2b and HiCal-2a were then launched, approximately 20 hours apart, in succession,
9 days after the initial ANITA-4 launch, with HiCal-2b leading ANITA and HiCal-2a trailing, each by several hundred
kilometers. The trajectories of the two HiCal payloads are shown in Figure \ref{fig:balloon1both}. As reported
elsewhere, both payloads were successfully tracked by a ground receiver array during ascent\citep{ShiHaoWang2017} with
2--3 degree precision in both azimuth and elevation during ascent.
In-flight
slow control parameters (temperature and pressure vessel pressure) for the HiCal-2a flight 
show a clear 24-hour cycle, consistent with the solar sky elevation and illumination.

\begin{figure}[htpb]\begin{floatrow}
\ffigbox[\FBwidth]{\includegraphics[width=0.5\textwidth]{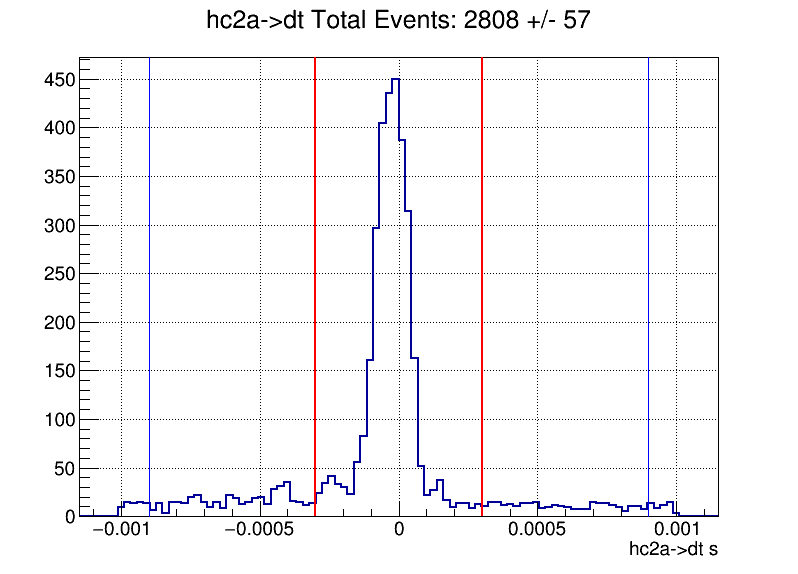}}{\caption{\it Time difference between a HiCal-a recorded trigger and the trigger time for recorded ANITA-4 events, corrected for signal propagation time. Red vertical lines indicate `signal' region; wider blue vertical lines designate 'sidebands' and are used to study ``background'' ANITA-4/HiCal-2 events having a `random' association only.}\label{fig:dtAHC2a}} 
\ffigbox[\FBwidth]{\includegraphics[width=0.5\textwidth]{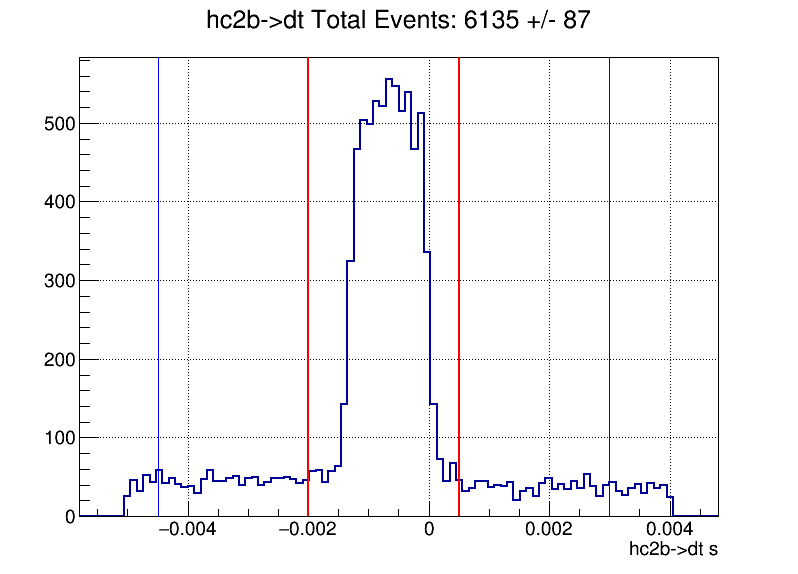}}{\caption{\it Time difference between a HiCal-b recorded trigger and the trigger time for recorded ANITA-4 events, corrected for signal propagation time.}\label{fig:dtAHC2b}} 
\end{floatrow}\end{figure}

There are several parameters that can be used to identify HiCal triggers in the ANITA-4 data sample. Most obviously,
we can compare the recorded HiCal transmitter trigger time to the receiver trigger times for ANITA-4 recorded events after correcting for the 
expected transit time between HiCal and ANITA (based on the known GPS locations of the two payloads) -- this
should yield a distribution that centers at zero, as indicated in Figures \ref{fig:dtAHC2a} and \ref{fig:dtAHC2b}. These
distributions readily identify HiCal events
for those transmitted pulses with telemetered timestamps.

\begin{figure}[htpb]\begin{floatrow}
\ffigbox[\FBwidth]{\includegraphics[width=0.5\textwidth]{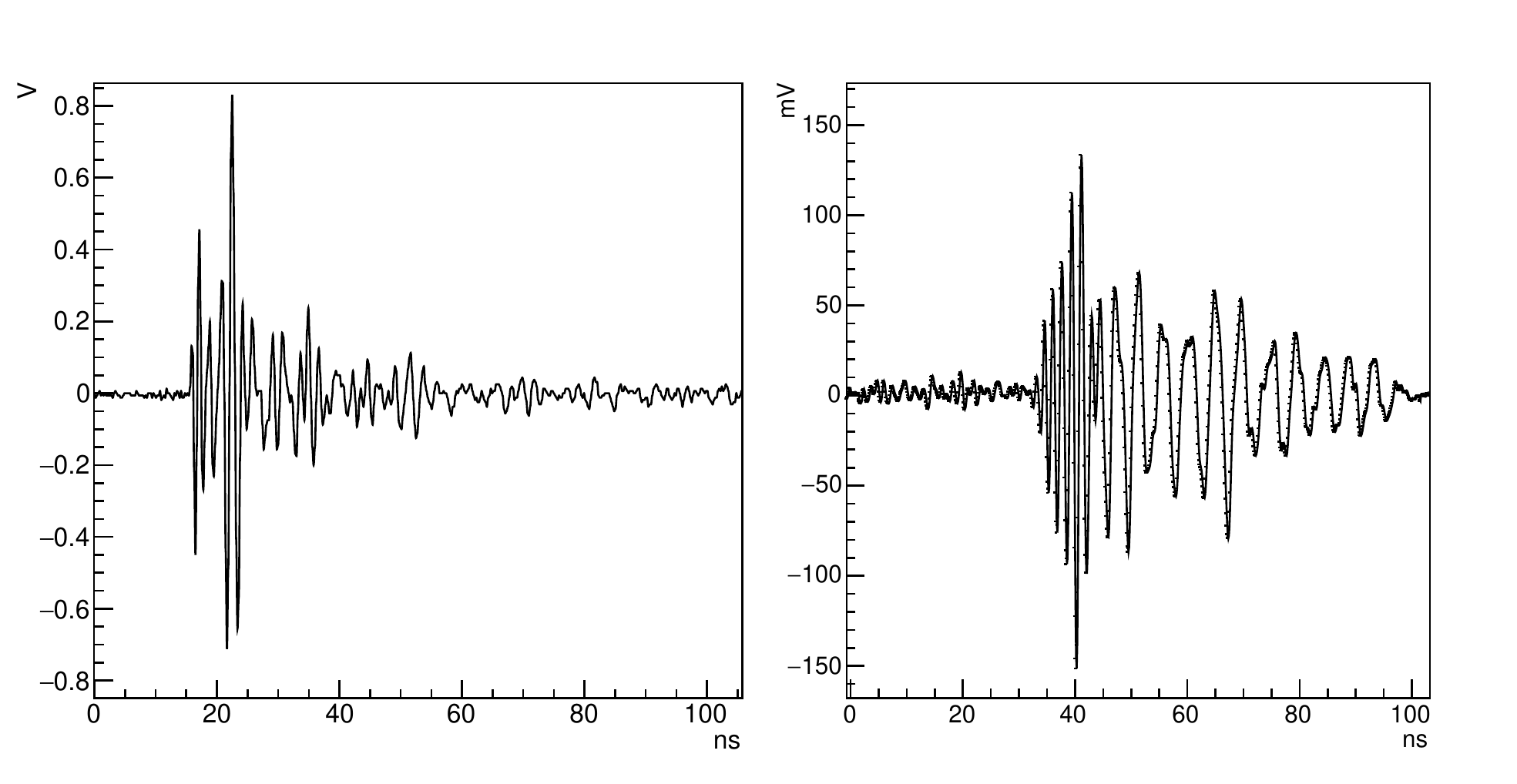}}
{\caption{\it Comparison of signals (Volts vs. nanoseconds) captured pre-flight (left) with actual ANITA-4 flight data (right). Horizontal scale units are nanoseconds. Signal tail evident at large times is largely an artifact of the ANITA-4 RF response.}\label{fig:compare_gnd_flight}}
\ffigbox[\FBwidth]{\includegraphics[width=0.5\textwidth]{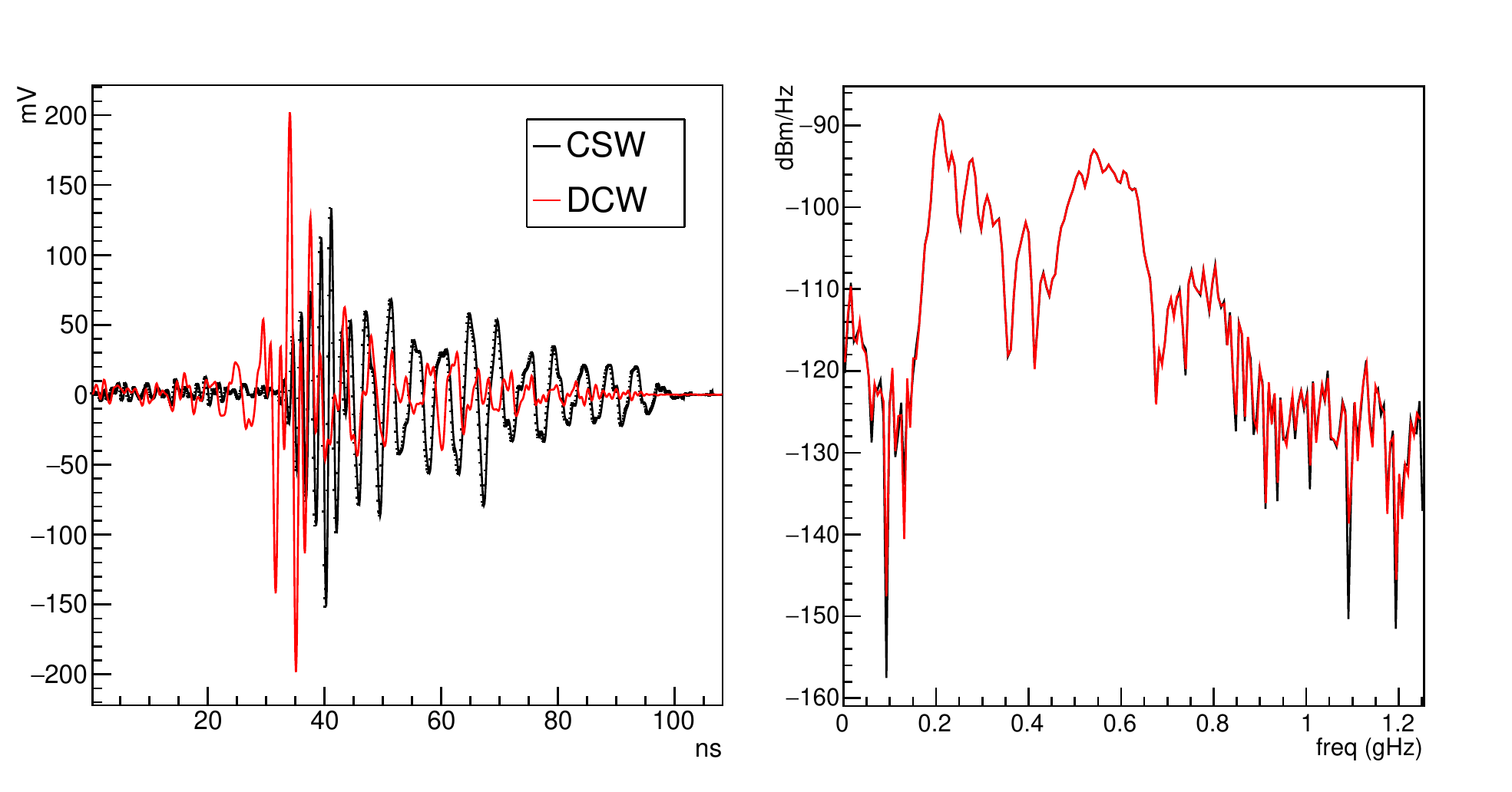}}
{\caption{\it HiCal received coherently summed waveform prior (black, left) and after (red, left) deconvolution of ANITA-4 signal response. As indicated by the similarity of the power spectra (right), the deconvolution process corrects for frequency-dependent phase delays in the ANITA-4 signal chain, but otherwise does not change the total power spectrum of measured signals.}\label{fig:deconvolve}}
\end{floatrow}
\end{figure}

The signals obtained in pre-flight testing in Palestine, TX, with 
the HiCal transmitter broadcasting to an ANITA-4 quad-ridged horn antenna, but read-out into a high-bandwidth
Tektronix digital scope, compared to triggers captured in-flight, are shown in Figure \ref{fig:compare_gnd_flight}.
The waveforms for the in-flight events have a characteristic low-frequency tail, which can be removed by deconvolution of the
system response in post-processing, as illustrated in Figure \ref{fig:deconvolve}. The deconvolution process is
necessary to infer the actual shape of the waveform reflected off the Antarctic surface. 

\section{Calculation: A General Treatment of Reflection of Spherical Waves}
We seek to compare our measurements with expectation for in-air signal reflecting off surface ice.
In our previous article\citep{gorham2017hical}, 
we presented initial numerical estimates of the reflectivity, as a function of
incidence angle at an interface
between two media with refractive indices $n$ and $n_1$. For our case,
these correspond to the refractive indices of air and ice, respectively. 
Our current treatment, following \citep{stratton2007electromagnetic}, comprises a decomposition of incident signal into a sum of plane
waves of different wave vectors.

We first consider
the case of a flat surface and then generalize to a sphere, neglecting the 
Earth's flattening at the Poles. For each plane wave, the reflected 
and transmitted waves are subject to the standard boundary conditions, 
from which we derive the standard reflection coefficients. 
After determining the electric and magnetic fields associated with each plane wave, integration over
all wave vectors gives the total field.

\begin{figure}[!ht]\begin{floatrow}
\ffigbox[\FBwidth]{\includegraphics[width=0.45\textwidth]{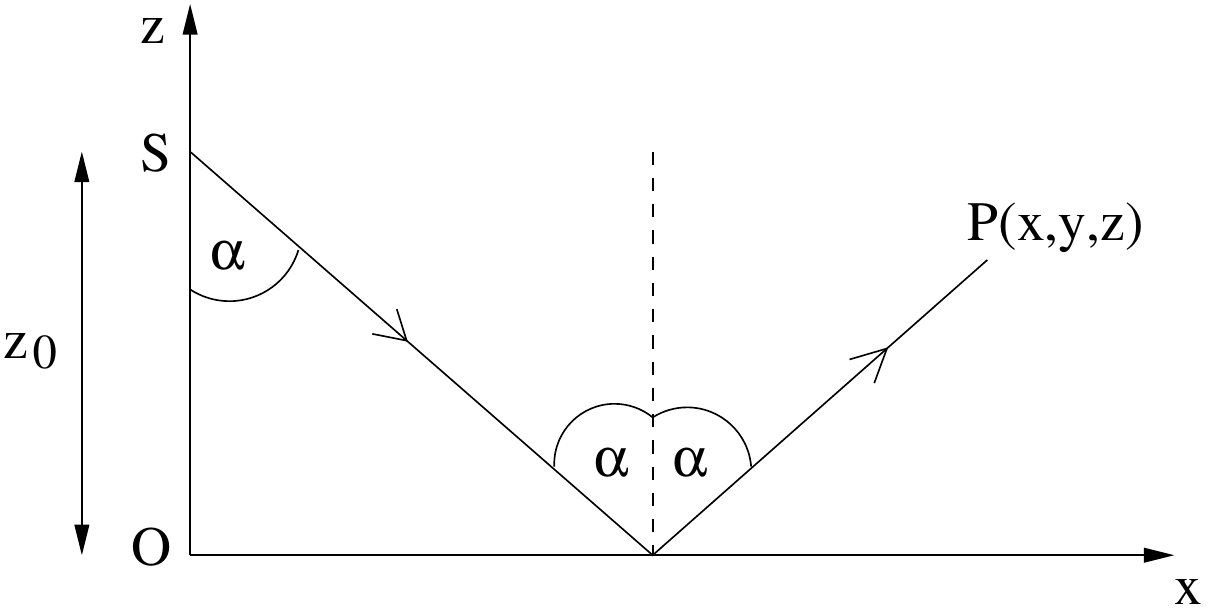}}
{\caption{\it Flat reflectivity calculation geometry: the source is located at $S$; $P$ represents any point with
position vector $\vec r=(x,y,z)$ with respect to the origin $O$.}\label{fig:geometry1}}
\ffigbox[\FBwidth]{\includegraphics[scale=0.45]{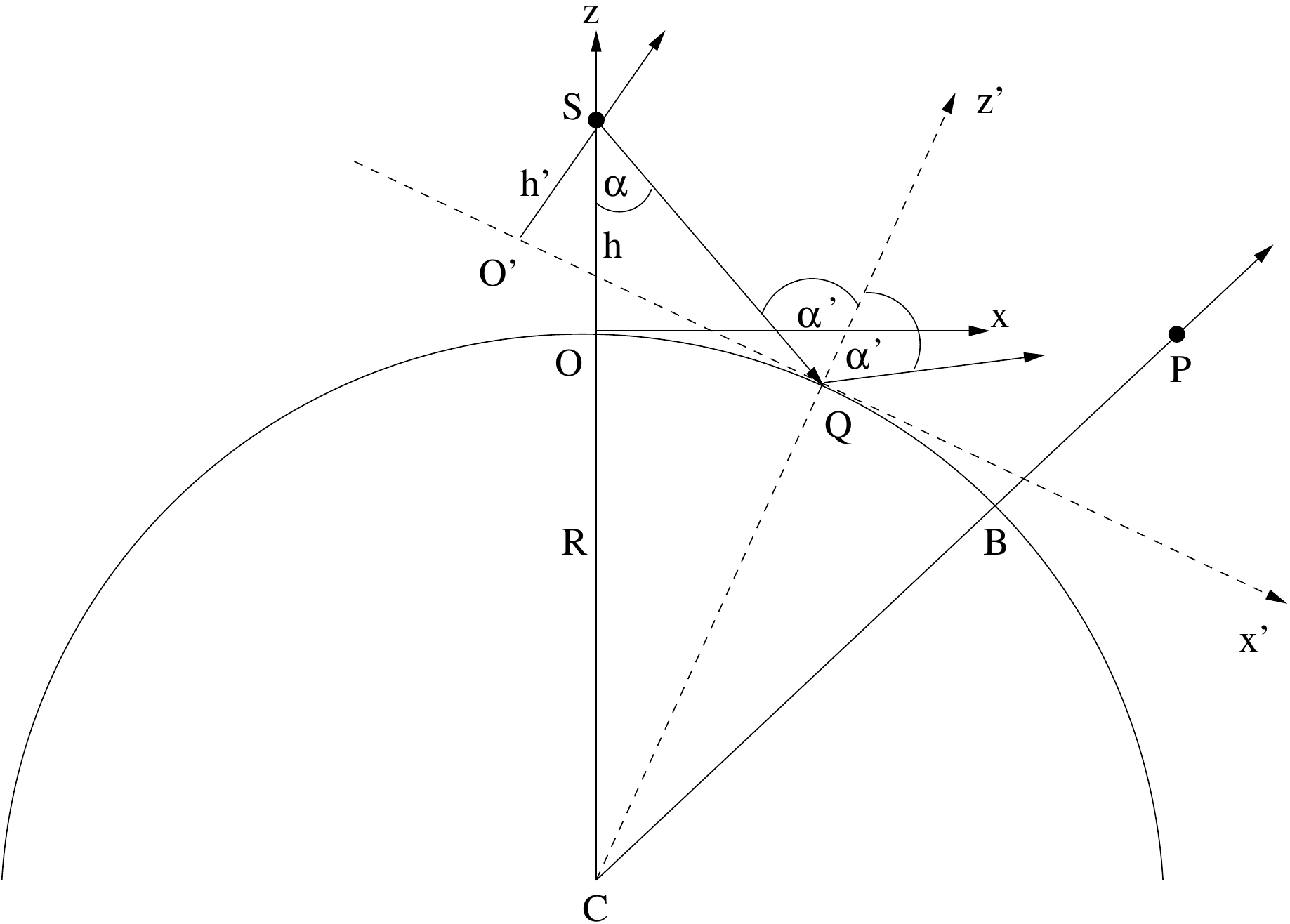}}
{\caption{\it Spherical reflectivity calculation geometry: The source is located at $S$,  the observer at $P$ and $O$ represents the origin of the coordinate system. For a particular plane wave corresponding to spherical coordinates $(\alpha,\beta)$ we identify the point $Q$ at which the wave vector originating from $S$ intersects the surface of the Earth. The corresponding reflected wave is assumed to be planar, and reflecting off the tangent plane at this point. For convenience, $\beta$ has been taken to be zero in this Figure. Here $O'$ refers to the origin of the transformed coordinate system and $\alpha'$ is the angle of reflection.}\label{fig:sph1}}
\end{floatrow}\end{figure}

The source is taken to be a dipole radiator, located at coordinates  $(0,0,z_0)$
and pointing towards the y-axis, i.e. with
a dipole moment $\hat p\propto \hat y$ as shown in Fig. \ref{fig:geometry1}. For comparison, the geometry for our subsequent calculation of 
the reflectivity for a spherical surface is shown in Fig. \ref{fig:sph1}. 
The configuration in which the dipole points along the $z$-axis has
been calculated in \citep{stratton2007electromagnetic} for the case of a flat surface. 
The observer
is located in the $x-z$ plane at $P(x,0,z)$. 
The surface of Earth, first assumed to be flat, coincides with the $x-y$ plane. 
The Hertz potential $\vec{\Pi}$
for such a radiator at position vector $\vec r=(x,y,z)$ 
for $z>0$ can be expressed as
\begin{equation}\label{eq:1}
\Pi_y(x,y,z) = {e^{ikR}\over 4\pi\epsilon R} + F_1(x,y,z)
\end{equation} 
with $\Pi_x=\Pi_z=0$ and $R = \sqrt{x^2+y^2+(z-z_0)^2}$. 
Here the first term on the right hand side represents
the primary radiation and the second term ($F_1$) arises due to 
reflection. 

For $0\le z\le z_0$,
the spherical wave can be decomposed as
\begin{equation}\label{eq:2}
{e^{ikR}\over R} = {ik\over 2\pi} \int_0^{2\pi}
\int_0^{{\pi\over 2}-i\infty} e^{ik[x\sin\alpha\cos\beta
+ y \sin\alpha\sin\beta + (z_0-z)\cos\alpha]} \sin\alpha d\alpha d\beta
\end{equation} 
with $\alpha$, and $\beta$ spherical polar coordinates. 
The right hand side represents an integral over plane waves. 
Note that the 1/R dependence on the 
left hand side of this expression is manifest in the oscillations of the exponential argument on the
right hand side, over which we integrate to determine the total signal strength at the final
observation point.
Alternately, we can interpret this equation as a superposition of plane waves, each
with a wave vector 
\begin{equation}
\vec k_I=k[\sin\alpha\cos\beta\hat x+ \sin\alpha\sin\beta\hat y
-\cos\alpha \hat z]\ ,
\end{equation}
i.e. with a polar angle $\pi-\alpha$ and azimuthal 
angle $\beta$. Note that, with this notation, we must integrate over complex
values of the polar angle. \message{Z: What does this mean?}
Using  
(\ref{eq:1}) and (\ref{eq:2}), we write the Hertz potential corresponding to an incident plane wave as  
\begin{equation}
\vec{\Pi}_{inc}=\frac{ik}{8\epsilon\pi^{2}}
\tilde\Pi
\hat{y}
\end{equation}
where
\begin{equation}
\tilde\Pi = 
e^{ikz_{0}\cos{\alpha}}
e^{ik(x\sin\alpha\cos\beta+y\sin\alpha\sin\beta-z\cos\alpha)} \,.
\end{equation}
The electric and magnetic fields can be computed using: 
\begin{eqnarray}
\vec E &=& \vec \nabla (\vec \nabla\cdot \vec \Pi) + k^2\vec \Pi\nonumber\\ 
\vec{H} &=& \frac{k^{2}}{i\omega\mu}(\vec\nabla\times\vec{\Pi})
\end{eqnarray}
where $\omega$ is the angular frequency of radiation and $\mu$ is the permeability of the medium. 

We are interested in the fields only in the Fraunh\"ofer far zone, $r>>\lambda$. 
The incident electric and magnetic fields are given by  
\begin{eqnarray}
 \vec{E}_{inc}&=& \frac{ik^{3}}{8\epsilon\pi^{2}}\tilde\Pi
\left[-\sin^{2}\alpha\cos\beta\sin\beta\hat{x}+(1-\sin^{2}\alpha\sin^{2}\beta)\hat{y}+(\sin\alpha\sin\beta\cos\alpha)\hat{z}\right]\nonumber\\
 \vec{H}_{inc}&=& \frac{ik^{2}\omega}{8\pi^{2}}\tilde\Pi
\left[\cos\alpha\hat{x}+(\cos\beta\sin\alpha)\hat{z}\right] \,.
\label{eq:incident}
\end{eqnarray}
To determine the reflected and transmitted fields, we first determine the plane of incidence for each plane wave. 
Next we decompose the electric and magnetic fields into independent components parallel
and perpendicular to the plane of incidence, then we integrate over the contributions from all the
plane waves. \\
\subsection{Reflection and Transmission on a Flat Surface} 
\label{eq:flat}{}
The basic geometry for a flat reflecting 
surface is illustrated in Fig. \ref{fig:geometry1}. To compare directly with HiCal, we focus on HPol.
For each incident
plane wave, we project the electric and magnetic field $\vec{E}$ and $\vec{H}$ into 
two components which are perpendicular and parallel
to the plane of incidence, i.e., 
\begin{equation}
 \vec{E}_{q}= \vec{E}^{s}_{q}+\vec{E}^{p}_q \nonumber
\end{equation}
\begin{equation} \label{eq:electric_1}
 \vec{H}_q= \vec{H}^{s}_q+\vec{H}^{p}_q
\end{equation}
where the subscript $q$ designates the incident, reflected or transmitted waves.
 For the electric field, $\perp$ and $\parallel$ components are denoted 
by the superscripts
$s$ and $p$, respectively. 
If the electric field lies in the plane of incidence then the corresponding
magnetic field is perpendicular to this plane. Hence for the case of
magnetic field, superscripts $s$ and $p$ denote components $\parallel$ and
$\perp$ to the plane of incidence, respectively. 
We next write the unit vector normal to the plane of incidence corresponding
to wave vector $\vec k_I$ as 
\begin{equation}
\hat \eta = l\hat{x}+m\hat{y}+n\hat{z}\,. 
\end{equation}
The vectors $\vec k_I$ and $\hat{z}$ lie in the plane of 
incidence and hence are perpendicular to 
$\hat\eta$. 
 This implies that $n=0$ and  
($l\hat{x}+m\hat{y}+n\hat{z})\cdot\vec k_I=0$. Hence we obtain 
$\hat \eta = (-\sin\beta\hat{x}+\cos\beta\hat{y})$. 
The vectors $\vec{E}^{s}_{q}$ and $\vec{H}^{p}_{q}$  
 point in the direction $\hat \eta$.
\\
 
For the incident wave, the $s$ and $p$ components of the electric field can be
expressed as:
\begin{equation}
 \vec{E}^{s}_{inc}=  \hat \eta [\vec{E}_{inc}\cdot \hat \eta]=\frac{ik^{3}}{8\epsilon\pi^{2}}\tilde\Pi
\, (-cos\beta\sin\beta\hat{x}+\cos^{2}\beta\hat{y})
\end{equation}
\begin{equation}\ 
\vec{E}^{p}_{inc}=  \vec{E}_{inc}-\vec{E}^{s}_{inc}= \frac{ik^{3}}
{8\epsilon\pi^{2}}\tilde \Pi
(\cos^{2}\alpha\cos\beta\sin\beta\hat{x}+\cos^{2}\alpha\sin^{2}\beta\hat{y}+\sin\alpha\cos\alpha\sin\beta\hat{z})
\label{eq:electric_2} \,.
\end{equation}
Similarly, the $s$ and $p$ components of the magnetic field are given by 
\begin{equation}
\vec{H}^{p}_{inc}=[\vec{H}_{inc}\cdot \hat \eta] 
\hat\eta=\frac{ik^{2}\omega}{8\pi^{2}}\tilde\Pi
(\cos\alpha\sin^{2}\beta\hat{x}-\cos\alpha\cos\beta\sin\beta\, \hat{y}) 
\end{equation}
\begin{equation}
\vec{H}^{s}_{inc}= \vec{H}_{inc}-\vec{H}^{p}_{inc} =  
\frac{ik^{2}\omega}{8\pi^{2}}\tilde\Pi
(\cos\alpha\cos^{2}\beta\hat{x}+\cos\alpha\cos\beta\sin\beta\hat{y}+\sin\alpha\cos\beta\hat{z}) \,.
\end{equation}
 For our case, we assume that the observer is located in the $x-z$ plane. In order to determine the reflected and transmitted waves we
 treat contributions from different $\vec k_I$ separately.
The $s$ component of the reflected wave is
straightforward. We obtain
\begin{equation}
 \vec{E}^{s}_{ref}
  =f^{s}_{r}\frac{ik^{3}}{8\epsilon\pi^{2}}\tilde \Pi_{ref}
(-\cos\beta\sin\beta\hat{x}+\cos^{2}\beta\hat{y})
\label{eq:erefs}
\end{equation}
where
\begin{equation}
\tilde\Pi_{ref} = 
e^{ikz_{0}\cos{\alpha}}
e^{ik(x\sin\alpha\cos\beta+y\sin\alpha\sin\beta+z\cos\alpha)} \,.
\end{equation}
\\
For the $p$ component we need to reverse the signs of the $x$ and $y$ 
components of Eq. (\ref{eq:electric_2}), leading to:
\begin{equation}
\vec{E}^{p}_{ref}= f^{p}_{r}\frac{ik^{3}}{8\epsilon\pi^{2}}\tilde\Pi_{ref}
(-\cos^{2}\alpha\cos\beta\sin\beta\hat{x}-\cos^{2}\alpha\sin^{2}\beta\hat{y}+\sin\alpha\cos\alpha\sin\beta\hat{z})
\label{eq:erefp}
\end{equation}
Similarly,  
\begin{equation}
\vec{H}^{p}_{ref}
=f^{p}_{r}\frac{ik^{2}\omega}{8\pi^{2}}\tilde\Pi_{ref}
(\cos\alpha\sin^{2}\beta\hat{x}-\cos\alpha\cos\beta\sin\beta\hat{y})
\end{equation}
and
\begin{equation}
\vec{H}^{s}_{ref}= f^{s}_{r}\frac{ik^{2}\omega}{8\pi^{2}}\tilde\Pi_{ref}
(-\cos\alpha\cos^{2}\beta\hat{x}-\cos\alpha\cos\beta\sin\beta\hat{y}+\sin\alpha\cos\beta\hat{z})\,.
\end{equation}
 where $f^{s}_{r}$  and $f^{p}_{r}$ are the reflection coefficients
 corresponding to the $s$ and $p$ components of the reflected fields, respectively. 

The corresponding transmitted fields $\vec{E}^{s}_{trans}$, $\vec{E}^{p}_{trans}$, $\vec{H}^{s}_{trans}$ and $\vec{H}^{p}_{trans}$ are obtained by the 
standard procedure. These have the same form as the incident wave with $k$ and
$\epsilon$ replaced by $k_1$ and $\epsilon_1$, respectively and are given by
\begin{equation} \label{eq:eflattrans}
\vec{E}^{s}_{trans}= f^{s}_{t}\frac{ik_{1}^{3}}{8\epsilon_{1}\pi^{2}}
(-\cos\beta_{t}\sin\beta_{t}\hat{x}+\cos^{2}\beta_{t}\hat{y})\tilde\Pi_t
\end{equation}
\begin{equation} \label{eq:eflattranp}
\vec{E}^{p}_{trans}=
f^{p}_{t}\frac{ik_{1}^{3}}{8\epsilon_{1}\pi^{2}}
(\cos^{2}\alpha_{t}\cos\beta_{t}\sin\beta_{t}\hat{x}+\cos^{2}\alpha_{t}\sin^{2}\beta_{t}\hat{y}+\cos\alpha_{t}\sin\alpha_{t}\sin\beta_{t}\hat{z})\tilde\Pi_t
\end{equation}
where
\begin{equation}
\tilde\Pi_t = 
e^{ikz_{0}\cos{\alpha}}e^{ik_{1}(x\sin\alpha_{t}\cos\beta_{t}+y\sin\alpha_{t}\sin\beta_{t}-z\cos\alpha_{t})}
\end{equation} 
\begin{equation}
\vec k_t=k_{1}[\sin\alpha_{t}\cos\beta_{t}\hat x+ \sin\alpha_{t}\sin\beta_{t}\hat y
-\cos\alpha_{t} \hat z]\ ,
\end{equation}
i.e. the transmitted wave vector $\vec k_t$ has a polar angle $\pi-\alpha_{t}$ and azimuthal angle $\beta_{t}$.\\
 
We point out that in the constant term, $e^{ikz_{0}\cos{\alpha}}$, in $\tilde\Pi_t$, 
the exponent is proportional to $k$ and not $k_1$. The overall
normalization of this term is contained in the reflection coefficients 
$f_t^s$ and $f_t^p$ which are fixed by the boundary conditions.  
The corresponding expressions for the transmited magnetic fields can be written as 
\begin{equation} 
\vec{H}^{p}_{trans}= f^{p}_{t}\frac{ik_{1}^{2}\omega}{8\pi^{2}}
(\cos\alpha_{t}\sin^{2}\beta_{t}\hat{x}-\cos\alpha_{t}\cos\beta_{t}\sin\beta_{t}\hat{y})\tilde\Pi_t\,,
\end{equation}
and
\begin{equation} 
 \vec{H}^{s}_{trans}=\vec{H}_{trans}-\vec{H}^{p}_{trans}= f^{s}_{t}\frac{ik_{1}^{2}\omega}{8\pi^{2}}
(\cos\alpha_{t}\cos^{2}\beta_{t}\hat{x}+\cos\alpha_{t}\cos\beta_{t}\sin\beta_{t}\hat{y}+\sin\alpha_{t}\cos\beta_{t}\hat{z})\tilde\Pi_t\,.
\end{equation}
\\
We next impose 
the boundary conditions at the $z=0$ interface on each component 
in order to determine
the reflection coefficients. The exponential factors lead to the standard 
conditions:
\begin{equation}
 k\sin\alpha=k_{1}\sin\alpha_{t}, \qquad \beta=\beta_{t}\,.
\end{equation}
The continuity of the electric field components parallel to the surface imply
that the $x$ and $y$ components are continuous, i.e., 
$$\vec{E}^{p}_{(trans),x,y}=\vec{E}^{p}_{(inc),x,y}+\vec{E}^{p}_{(ref),x,y}\,.$$
The perpendicular components follow:
 $$\epsilon_{1}\vec{E}^{p}_{trans,z}=\epsilon[\vec{E}^{p}_{inc,z}+
\vec{E}^{p}_{ref,z}]\,.$$
The perpendicular components of the magnetic field are continuous at
the interface and the parallel components satisfy
 $$\mu_{1}\vec{H}^{p}_{(trans),x,y}=\mu\left[\vec{H}^{p}_{(inc),x ,y}+
\vec{H}^{p}_{(ref),x,y}\right]\,.$$ 
Here we shall assume $\mu_{1}=\mu$.
These conditions lead to:
\begin{equation} 
(1-f^{p}_{r})= f^{p}_{t}\frac{k_{1}}{k}\frac{\cos^{2}\alpha_{t}}{\cos^{2}\alpha}
\label{eq:Fresnelp1}
\end{equation}
\begin{equation}
 (1+f^{p}_{r})= f^{p}_{t}\frac{k^{3}_{1}}{k^{3}}\frac{\cos\alpha_{t}\sin\alpha_{t}}{\cos\alpha\sin\alpha}
 = f^{p}_{t}\frac{k^{2}_{1}}{k^{2}}\frac{\cos\alpha_{t}}{\cos\alpha}\,.
\label{eq:Fresnelp2}
\end{equation}
 Solving Eqs. \ref{eq:Fresnelp1} and \ref{eq:Fresnelp2} we obtain 
\begin{equation}
f^{p}_{r} = \frac{k_{1}\cos\alpha-k\cos\alpha_{t}}{k_{1}\cos\alpha+k\cos\alpha_{t}} 
\end{equation}
and 
\begin{equation}
f^{p}_{t}= \left(\frac{k}{k_{1}}\right)^{2}\left(\frac{1}{\cos\alpha_{t}}
\right)\frac{2k_{1}\cos^{2}\alpha}{k_{1}\cos\alpha+k\cos\alpha_{t}}\,.
\end{equation}
\\
 We next impose boundary conditions on the components perpendicular
to the plane of incidence. These lead to 
 $$\vec{E}^{s}_{(trans),x ,y}=\vec{E}^{s}_{(inc),x ,y}+\vec{E}^{s}_{(ref),x ,y}$$
 and
  $$\vec{H}^{s}_{(trans),x,y}=\vec{H}^{s}_{(inc),x,y}+\vec{H}^{s}_{(ref),x,y}\,.$$
These conditions imply
\begin{equation}
 (1+f^{s}_{r})= f^{s}_{t}\frac{k_{1}}{k}
 \label{eq:4s}
\end{equation}
and
\begin{equation}
 (1-f^{s}_{r})= f^{s}_{t}\frac{k^{2}_{1}}{k^{2}}\frac{\cos\alpha_{t}}{\cos\alpha}\,.
 \label{eq:5s}
\end{equation}
 Solving Eqs. \ref{eq:4s} and \ref{eq:5s} we obtain:
\begin{equation}
f^{s}_{r} = \frac{k\cos\alpha-k_{1}\cos\alpha_{t}}{k\cos\alpha+k_{1}\cos\alpha_{t}}
\end{equation}
and
\begin{equation}
f^{s}_{t}= \left(\frac{k}{k_{1}}\right)^{2}\frac{2k_{1}\cos\alpha}{k_{1}\cos\alpha_{t}+
k\cos\alpha}\,.
\end{equation}
\\
Using the above reflection coefficients we can compute the $s$ and $p$ components of the reflected and 
transmitted fields for each plane wave. Adding Eqs. \ref{eq:erefs} and \ref{eq:erefp}, we find the total 
reflected electric field for each plane wave. 
Since we are interested only in 
the perpendicular component, we consider only the $y$-component of the
reflected field, obtained by integrating over the angles $\alpha$ and $\beta$, as\\
\begin{equation}
E_{(ref),y}={ik^{3}\over 8\epsilon\pi^{2}} \int_0^{2\pi}
\int_0^{{\pi\over 2}-i\infty} \tilde\Pi_{ref}(f^{s}_{r}\cos^{2}\beta-f^{p}_{r}\cos^{2}\alpha\sin^{2}\beta) \sin\alpha d\alpha d\beta\,.
\end{equation} 
\\
Similarly, we add Eqs. \ref{eq:eflattrans} and \ref{eq:eflattranp} to get the transmitted electric field for each plane wave. The final expression for the $y$-component of transmitted electric field is given by
\begin{equation}
E_{(trans),y}={ik_{1}^{3}\over 8\epsilon_{1}\pi^{2}} \int_0^{2\pi}
\int_0^{{\pi\over 2}-i\infty} \tilde\Pi_{t}(f^{s}_{t}\cos^{2}\beta_{t}+f^{p}_{t}\cos^{2}\alpha_{t}\sin^{2}\beta_{t}) \sin\alpha d\alpha d\beta \,.
\label{eq:totalflat}
\end{equation} 
We compute the reflection coefficient numerically as a function of the specular angle by setting the altitude of both the source and observer to
be 37 km, appropriate for HiCal-2 broadcasting to ANITA-4 at float altitude. For proper comparison we set the distance of propagation of the incident wave to be same as that of the reflected wave. The resulting value
of the reflection coefficient is found to be same as that for Fresnel reflection independent of frequency.
%
\subsection{Reflection and Transmission at a Spherical Surface} 
\label{eq:spherical}
In this section we derive the reflection coefficient for a spherical interface between 
air and ice. The source $S$ is again assumed to be a dipole located at
an altitude of $h$. In a Cartesian coordinate system centered at $O$, the transmitter
coordinates are $(0,0,z_0)$ with $z_0=h$ (see Fig. \ref{fig:sph1}). 
As in the case of a flat surface, we again decompose the spherical
wave in terms of plane waves. 
In contrast to the case of a flat surface, the reflected wave corresponding
to each incident plane wave will not be a plane wave. However since the curvature
is small it may be reasonable to approximate it as a plane wave. This is
justified by the observation, as discussed in more detail later,
 that the dominant contribution to
the reflected wave arises from a small angular region near the
specular point.   
For each plane wave 
corresponding to spherical polar angles $(\alpha,\beta)$, we identify a point $Q$ 
on the spherical surface where the wave vector from the source $S$ intersects
the surface (see Fig. \ref{fig:sph1} ). 
We next assume that the reflection and refraction occurs on the plane tangent to $Q$.

For each incident plane wave, we transform our coordinate system
such that the new axes ($x',y'$) lie on the tangent plane and
the plane of reflection is same as the $x'-z'$ plane. We can now use our
planar reflection coefficients in this new coordinate system $(x^{\prime}-y^{\prime}-z^{\prime})$. First, we compute the electric and magnetic field components for each plane wave in this coordinate system.  
As the primed coordinate system is not fixed, and depends on the point of reflection Q, we transform
back to the original frame and integrate over all plane
waves to get the total field.    
\\
For a given plane wave, let the point $Q$ be located at 
$(x_{s},y_{s},z_{s})$. We identify the tangent plane at this point 
 and choose the cordinate system $(x^{\prime}-y^{\prime}-z^{\prime})$ 
such that it satisfies the following conditions:
\begin{itemize} 
\label{eq:condition}{}
\item[1.] The coordinates of $Q$ in this new coordinate system are
 $(x^{\prime}_{s},0,0)$. 
\item[2.]  The source point $S$ 
in the new coordinate system lies at $(0,0,h^{\prime})$.  
\item[3.]  The unit vector normal to the tangent plane at 
$Q$ is parallel to the $z^{\prime}$ axis.
\end{itemize}
This is accomplished by two rotations followed by a translation. We 
first rotate our coordinate system counter-clockwise
 about the $z$ axis by an angle $\beta$. The rotation matrix corresponding
to this is  
\begin{equation}
 R_{z}(\beta)=\left(\begin{array}{ccc}\cos\beta & \sin\beta & 0 \\ -\sin\beta & \cos\beta & 0 \\ 0 & 0 & 1 \end{array}\right)\,.
\end{equation}
Next we rotate counter-clockwise about the new $y$-axis by an angle  
 ($\alpha^{\prime}-\alpha$). This leads to the rotation matrix 
\begin{equation}
R_{y}(\alpha^{\prime}-\alpha)= \left(\begin{array}{ccc} \cos(\alpha'-\alpha) & 0 & -\sin(\alpha'-\alpha)\\ 0 & 1 & 0 \\ \sin(\alpha'-\alpha) & 0 & \cos(\alpha'-\alpha) \end{array}\right)\,.
\end{equation}
Now the overall rotation matrix is given by, \
 $Rot=  R_{y}(\alpha^{\prime}-\alpha) R_{z}(\beta)$\
\begin{equation} 
Rot = \left(\begin{array}{ccc}\cos(\alpha'-\alpha)\cos\beta & \cos(\alpha'-
\alpha)\sin\beta & -\sin(\alpha'-\alpha)\\ -\sin\beta & \cos\beta & 0 \\ 
\sin(\alpha'-\alpha)\cos\beta &\sin(\alpha'-\alpha)\sin\beta & \cos(\alpha'-
\alpha) \end{array}\right)\,.
\label{eq:Rot}
\end{equation}
With these two rotations, we obtain the 
coordinate system $(x''-y''-z'')$ which satisfies condition
3. given above and the tangent plane becomes parallel to the $x''-y''$ plane. 
We next apply a translation in the $(x''-y''-z'')$ coordinate system given by:
\begin{eqnarray}
 x''_{0}&=& -h\sin(\alpha^{\prime}-\alpha),\nonumber\\ 
 y''_{0}&=&  0 ,\nonumber\\
 z''_{0}&=& \frac{1}{2}\left(R+2h\cos(\alpha^{\prime}-\alpha)-\frac{R\sin(2\alpha^{\prime}-\alpha)}{\sin\alpha}\right)\,.
\end{eqnarray} \\
This leads to the final coordinate system $x'-y'-z'$ which
satisfies all the conditions 
given above and
has the origin located at $O'$. 
 The angle $\alpha'$ is the angle of reflection as shown
in Fig. \ref{fig:sph1}. \\
For each incident plane wave we can now use the 
 formalism developed in section \ref{eq:flat} for a flat surface. 
 We obtain the coordinates of the observation point P in the new  
system $(x^{\prime},y^{\prime},z^{\prime})$ by applying the Rotation above 
followed by a translation in the $x''-y''-z''$ frame. The 
observation point $P(x^{\prime},y^{\prime},z^{\prime})$ in the new coordinate system is given by:
 \begin{equation}
 \left(\begin{array}{ccc} x^{\prime}\\  y^{\prime}\\ z^{\prime}\end{array}\right)= Rot \cdot \left(\begin{array}{ccc} x\\  y\\ z \end{array}\right)-\left(\begin{array}{ccc} x''_{0}\\  y''_{0}\\ z''_{0} \end{array}\right)\,.
\end{equation} 
\\
We now find the incident, reflected and transmitted fields for the spherical geometry defined in Fig. \ref{fig:sph1}. 
The exponent appearing in the expression for $\vec{\Pi}$ in Section \ref{eq:flat} is now dependent on the geometry of reflecting surface, coordinates of point of observation and 
the dipole height in the new frame. The basis vectors in this new coordinate system are related to those in the 
old coordinate system by the formulae
\begin{eqnarray}
\hat{x}^\prime &=& \cos(\alpha'-\alpha)(\cos\beta\hat x + \sin\beta\hat y)
- \sin(\alpha'-\alpha)\hat z\nonumber\\ 
\hat{y}^\prime &=& -\sin\beta\hat x + \cos\beta\hat y \nonumber\\ 
\hat{z}^\prime &=& \sin(\alpha'-\alpha)(\cos\beta\hat x + \sin\beta\hat y)
+ \cos(\alpha'-\alpha)\hat z \,.
\end{eqnarray}
We next write the incident wave vector in the new coordinate system as:
\begin{equation}
 \vec{k}^{\prime}_{inc}= Rot \cdot \vec{k}_{inc} = k(\sin\alpha^{\prime}\hat{x}^{\prime}-\cos\alpha^{\prime}\hat{z}^{\prime})\,.
 \end{equation}
The reflected wave vector in the new frame is given by:
\begin{equation}
 \vec{k}^{\prime}_{ref}= Rot \cdot \vec{k}_{ref} = k(\sin\alpha^{\prime}\hat{x}^{\prime}+\cos\alpha^{\prime}\hat{z}^{\prime})\,.
 \end{equation}
 We write the corresponding transmitted wave vector $\vec{k}^{\prime}_{trans}$ as:
 \begin{equation}
 \vec{k}^{\prime}_{trans}= Rot \cdot \vec{k}_{trans} = k_{1}(\sin\alpha_{t}^{\prime}\hat{x}^{\prime}-\cos\alpha_{t}^{\prime}\hat{z}^{\prime})
 \end{equation}
where $\pi-\alpha_{t}^{\prime}$ and $\beta_{t}^{\prime}=\beta_t$ are, 
respectively, the polar and azimuthal angles of  $\vec{k}^{\prime}_{trans}$.
The exponential factor for the incident plane wave is derived for spherical geometry using the same method as in the case of flat geometry. We express it as
\begin{equation}
\tilde\Pi_{S,i} = \exp\left[{ik'_{inc}\cdot (\vec r^{\,\prime} -h'\hat{z}^\prime) }\right]
\end{equation} 
where (0, 0, $h'$) is the location of the dipole in the new 
frame and the point of observation is located at 
vector $\vec{r}^{\,\prime}$ with respect to the new origin $O^{\prime}$. 
The exponential factor for the reflected plane wave is obtained from geometry (Fig. \ref{fig:sph1}) as
\begin{equation}
\tilde\Pi_{S,r} = \exp\left[{ik'_{ref}\cdot (\vec r^{\,\prime} +h'\hat{z}^\prime) }\right]\,.
\end{equation} 
In the transformed frame we may again identify the location of the image
as in the case of flat geometry \citep{stratton2007electromagnetic}. Let the image be located
at the position vector $\vec \Delta$ with respect to the origin of
the original coordinate system. We then have 
\begin{equation}
\tilde\Pi_{S,r} = \exp\left[{ik'_{ref}\cdot (\vec r -\vec \Delta) }\right]\,.
\end{equation} 
We obtain
\begin{equation}
\vec k'_{ref}\cdot \vec r = k\left[x\sin(2\alpha'-\alpha)\cos\beta
+ y \sin(2\alpha'-\alpha)\sin\beta + z\cos(2\alpha'-\alpha)  \right]
\end{equation} 
and
\begin{equation}
 k'_{ref}\cdot \vec \Delta =  k\left[z_0\cos(2\alpha'-\alpha) -
2h'\cos\alpha'\right]
\end{equation} 
where 
\begin{equation}
h' = R{\sin(\alpha'-\alpha)\cos\alpha'\over \sin\alpha}
\end{equation} 
and
\begin{equation}
 \sin\alpha^{\prime}= \frac{(R+h)\sin\alpha}{R}
\end{equation}
This can be derived easily by using geometry.
We see from Fig. \ref{fig:sph1} that the exponential factor can also be written as:
$$\exp\left({i[\vec{k}^{\prime}_{inc}\cdot (\vec{r}^{\,\prime}_{s}-\vec{h}^{\prime})+\vec{k}^{\prime}_{ref}\cdot (\vec{r}^{\,\prime}-\vec{r}^{\,\prime}_{s})]}
\right)$$ 
where
\begin{equation}
\vec{r}^{\,\prime}_{s}= \frac{R\sin(\alpha^{\prime}-\alpha)\sin\alpha^{\prime}}{\sin\alpha}\hat{x}^{\prime}, \nonumber
\end{equation}
\begin{equation}
\vec{r}^{\,\prime}= x^{\prime}\hat{x}^{\prime}+y^{\prime}\hat{y}^{\prime}+z^{\prime}\hat{z}^{\prime}  \nonumber
\end{equation}
and 
\begin{equation}
\vec h^{\prime}= h^\prime\hat{z}^{\prime}\,.
\end{equation}
This provides an alternative
way to derive the formula for the exponent appearing in $\tilde\Pi_{S,r}$ and yields the same result as 
before. 
The exponential factor for the transmitted wave obtains from geometry 
(see Fig. \ref{fig:sph1}), and can be expressed as
\begin{equation}
\tilde\Pi_{S,t} = e^{i\vec{k}^{\prime}_{inc}\cdot\vec{\Delta}^{\prime}}e^{i\vec{k}^{\prime}_{trans}\cdot \vec r^{\,\prime}}
\end{equation} 
where $e^{i\vec{k}^{\prime}_{inc}\cdot\vec{\Delta}^{\prime}}$ is the constant term appearing in both $\tilde\Pi_{S,i}$ and $\tilde\Pi_{S,t}$. As in the case of flat geometry, this term is proportional to k and not $k_{1}$. 

We next write the electric and magnetic field components in the new coordinate system:
\begin{equation}  
 \vec{E}^{\prime}_{inc}= Rot \cdot \vec{E}_{inc} \nonumber
 \end{equation}
\begin{equation}\label{eq:prime}
 \vec{H}^{\prime}_{inc}= Rot \cdot \vec{H}_{inc} \,.
 \end{equation}
Using Eqs. \ref{eq:incident} and  \ref{eq:Rot}  we obtain
\begin{equation}
\vec{E}^{\prime}_{inc}= 
\frac{ik^3}{8\epsilon\pi^{2}}\,
\tilde \Pi_{S,i} \, [\cos\alpha\cos\alpha^{\prime}\sin\beta\hat{x}^{\prime}+\cos\beta\hat{y}^{\prime}+\cos\alpha\sin\alpha^{\prime}\sin\beta\hat{z}^{\prime}]
\end{equation}
and 
\begin{equation}
\vec{H}^{\prime}_{inc}= 
\frac{ik^2\omega}{8\pi^{2}}\,
\tilde \Pi_{S,i} \, [\cos\alpha^{\prime}\cos\beta\hat{x}^{\prime}-\cos\alpha\sin\beta\hat{y}^{\prime}+\sin\alpha^{\prime}\cos\beta\hat{z}^{\prime}]\,.
\end{equation}
Now we use the same method as in the case of flat geometry to find the $s$ 
and
 $p$ components of $E^{\prime}_{q}$ and  $H^{\prime}_{q}$ (where $q$ again denotes the incident, reflected or transmitted wave) and use the boundary conditions at $z^\prime_{s}=0$ to find the reflection coefficients.

 In order to calculate the s and p components of the electric and magnetic fields, we first need to find a unit vector which is 
 perpendicular to both $\vec{k}^{\prime}_{inc}$ and
$\hat{z}^{\prime}$. 
The resulting unit
 vector $\hat{l}$ perpendicular to the plane of incidence is given by 
\begin{equation}
 \hat{l}= \hat{y}^{\prime}\,.
\end{equation}
Now we can write the incident electric field components as
\begin{equation}
\vec{E}^{\prime(s)}_{inc}= (\vec{E}^{\prime}_{inc}\cdot \hat{l})\hat{l}=   
\frac{ik^3}{8\epsilon\pi^{2}}\,
\tilde \Pi_{S,i}\, [\cos\beta \hat{y}^{\prime}]\nonumber
\end{equation}
\begin{equation}
\vec{E}^{\prime(p)}_{inc}= \vec{E}^{\prime}_{inc}-\vec{E}^{\prime(s)}_{inc}=\frac{ik^3}{8\epsilon\pi^{2}}\,
\tilde \Pi_{S,i}\,
[\cos\alpha\cos\alpha^{\prime}\sin\beta \hat{x}^{\prime}+\cos\alpha\sin\alpha^{\prime}\sin\beta \hat{z}^{\prime}]\,.
 \end{equation}
Similarly the incident magnetic field components can be written as 
\begin{equation}
\vec{H}^{\prime(p)}_{inc}= (\vec{H}^{\prime}_{inc}\cdot \hat{l})\hat{l}=\frac{ik^2\omega}{8\pi^{2}}\,
\tilde \Pi_{S,i}\, [-\cos\alpha\sin\beta \hat{y}^{\prime}] \nonumber
\end{equation}
\begin{equation}
 \vec{H}^{\prime(s)}_{inc}= \vec{H}^{\prime}_{inc}-\vec{H}^{\prime(p)}_{inc}=\frac{ik^2\omega}{8\pi^{2}}\,
\tilde \Pi_{S,i}\,
[\cos\alpha^{\prime}\cos\beta \hat{x}^{\prime}+\sin\alpha^{\prime}\cos\beta \hat{z}^{\prime}]\,.
 \end{equation}
 The $s$ and $p$ components of the reflected electric field are obtained as
 \begin{equation}
  \vec{E}^{\prime(s)}_{ref}=  
f^{\prime (s)}_{r}\frac{ik^3}{8\epsilon\pi^{2}}\,
\tilde \Pi_{S,r}\, [\cos\beta \hat{y}^{\prime}] \nonumber,
 \end{equation}

\begin{equation}
 \vec{E}^{\prime(p)}_{ref}= f^{\prime (p)}_{r} \frac{ik^3}{8\epsilon\pi^{2}}\,
\tilde \Pi_{S,r}\,
[-\cos\alpha\cos\alpha^{\prime}\sin\beta \hat{x}^{\prime}+\cos\alpha\sin\alpha^{\prime}\sin\beta \hat{z}^{\prime}]\,.
 \end{equation}
Similarly, for the reflected magnetic field components we write
\begin{equation}
  \vec{H}^{\prime(p)}_{ref}= f^{\prime (p)}_{r}  \frac{ik^2\omega}{8\pi^{2}}\,
\tilde \Pi_{S,r}\, [-\cos\alpha\sin\beta \hat{y}^{\prime}]
 \end{equation}
\begin{equation}
\vec{H}^{\prime(s)}_{ref}= f^{\prime (s)}_{r} \frac{ik^2\omega}{8\pi^{2}}\,
\tilde \Pi_{S,r}\,
[-\cos\alpha^{\prime}\cos\beta \hat{x}^{\prime}+\sin\alpha^{\prime}\cos\beta \hat{z}^{\prime}]
\end{equation} 
 where $f^{\prime (s)}_{r}$ and $f^{\prime (p)}_{r}$ are the reflection coefficients corresponding to the 
$s$ and $p$ components of the reflected fields. 
\\ 
The corresponding transmitted fields $\vec{E}^{\prime(s)}_{trans}$, $\vec{E}^{\prime(p)}_{trans}$, $\vec{H}^{\prime(s)}_{trans}$ and $\vec{H}^{\prime(p)}_{trans}$ can be written as
\begin{equation}
\vec{E}^{\prime(s)}_{trans}=    f^{\prime (s)}_{t} \frac{ik_{1}^3}{8\epsilon_{1}\pi^{2}}\,
\tilde \Pi_{S,t}\, [\cos\beta_{t} \hat{y}^{\prime}]\nonumber
\label{eq:etrans}
\end{equation}
\begin{equation}
\vec{E}^{\prime(p)}_{trans}= \vec{E}^{\prime}_{trans}-\vec{E}^{\prime(s)}_{trans}=f^{\prime (p)}_{t}\frac{ik_{1}^3}{8\epsilon_{1}\pi^{2}}\,
\tilde \Pi_{S,t}\,
[\cos\alpha_{t}\cos\alpha_{t}^{\prime}\sin\beta_{t} \hat{x}^{\prime}+\cos\alpha_{t}\sin\alpha_{t}^{\prime}\sin\beta_{t} \hat{z}^{\prime}]\,.
\label{eq:etranp}
 \end{equation}
Similarly the transmitted magnetic field components can be written as 
\begin{equation}
\vec{H}^{\prime(p)}_{trans}= f^{\prime (p)}_{t}\frac{ik_{1}^2\omega}{8\pi^{2}}\,
\tilde \Pi_{S,t}\, [-\cos\alpha_{t}\sin\beta_{t} \hat{y}^{\prime}] \nonumber
\end{equation}
\begin{equation}
 \vec{H}^{\prime(s)}_{trans}= \vec{H}^{\prime}_{trans}-\vec{H}^{\prime(p)}_{trans}=f^{\prime (s)}_{t}\frac{ik_{1}^2\omega}{8\pi^{2}}\,
\tilde \Pi_{S,t}\,
[\cos\alpha_{t}^{\prime}\cos\beta_{t} \hat{x}^{\prime}+\sin\alpha_{t}^{\prime}\cos\beta_{t} \hat{z}^{\prime}]\,.
 \end{equation}
\\
 We impose the boundary conditions at $z^{\prime}_{s}=0$ on each component in order to determine the reflection coefficients for reflection and transmission of a plane wave, using the same procedure described in section \ref{eq:flat}. The exponential factors lead to the standard conditions:
\begin{equation}
 k\sin\alpha^{\prime}=k_{1}\sin\alpha_{t}^{\prime} , \qquad \beta_{t}= \beta\,.
\end{equation}
The continuity of electric field components parallel to the surface imply that 
$$\vec{E}^{\prime p}_{trans,x}=\vec{E}^{\prime p}_{inc,x}+\vec{E}^{\prime p}_{ref,x}\,.$$
The components perpendicular to the surface follow:
 $$\epsilon_{1}\vec{E}^{\prime p}_{trans,z}=\epsilon[\vec{E}^{\prime p}_{inc,z}+
\vec{E}^{\prime p}_{ref,z}]\,.$$
The component of magnetic field $\perp$ to the surface are continuous at
the interface and the parallel components satisfy
 $$\mu_{1}\vec{H}^{\prime p}_{trans,y}=\mu\left[\vec{H}^{\prime p}_{inc,y}+
\vec{H}^{\prime p}_{ref,y}\right]\,.$$ 
Here we shall assume $\mu_{1}=\mu$.
These conditions lead to:
\begin{equation}
  (1-f^{\prime p }_{r})= f^{\prime p }_{t}\frac{k_{1}}{k}\frac{\cos\alpha_{t}\cos\alpha^{\prime}_{t}}{\cos\alpha\cos\alpha^{\prime}}
\label{eq:Fresnelp11}
\end{equation}
\begin{equation}
 (1+f^{\prime p }_{r})= f^{\prime p }_{t}\frac{k^{3}_{1}}{k^{3}}\frac{\cos\alpha_{t}\sin\alpha^{\prime}_{t}}{\cos\alpha\sin\alpha^{\prime}}
 =  f^{\prime p }_{t}\frac{k^{2}_{1}}{k^{2}}\frac{\cos\alpha_{t}}{\cos\alpha}\,.
\label{eq:Fresnelp22}  
\end{equation}
 Solving Eqs. \ref{eq:Fresnelp11} and \ref{eq:Fresnelp22} we obtain 
\begin{equation}
f^{\prime p}_{r} = \frac{k_{1}\cos\alpha^{\prime}-k\cos\alpha^{\prime}_{t}}{k_{1}\cos\alpha^{\prime}+k\cos\alpha^{\prime}_{t}} 
\end{equation}
and 
\begin{equation}
f^{\prime p}_{t}= \left(\frac{k}{k_{1}}\right)^{2}\left(\frac{1}{\cos\alpha_{t}}
\right)\frac{2k_{1}\cos\alpha\cos\alpha^{\prime}}{k_{1}\cos\alpha^{\prime}+k\cos\alpha^{\prime}_{t}}\,.
\end{equation}
We next impose boundary conditions on the components $\perp$ to the plane of incidence. These lead to
$$\vec{E}^{\prime s}_{trans,y}=\vec{E}^{\prime s}_{inc,y}+\vec{E}^{\prime s}_{ref,y}$$
 and
 $$\mu_{1}\vec{H}^{\prime p}_{trans,x}=\mu\left[\vec{H}^{\prime p}_{inc,x }+
\vec{H}^{\prime p}_{ref,x}\right]\,.$$
These conditions imply
\begin{equation}
 (1+f^{\prime s}_{r})= f^{\prime s}_{t}\frac{k_{1}}{k}
 \label{eq:4}
\end{equation}
and
\begin{equation}
 (1-f^{\prime s}_{r})= f^{\prime s}_{t}\frac{k^{2}_{1}}{k^{2}}\frac{\cos\alpha^{\prime}_{t}}{\cos\alpha^{\prime}}\,.
 \label{eq:5}
\end{equation}
 Solving Eqs. \ref{eq:4} and \ref{eq:5} we obtain:
 \begin{equation}
  f^{\prime s }_{r}= \frac{k\cos\alpha^{\prime}-k_{1}\cos\alpha_{t}^{\prime}}{k\cos\alpha^{\prime}+k_{1}\cos\alpha_{t}^{\prime}} \nonumber ,
  \end{equation}
and
\begin{equation}
f^{\prime s}_{t}= \left(\frac{k}{k_{1}}\right)^{2}\frac{2k_{1}\cos\alpha^{\prime}}{k\cos\alpha^{\prime}+k_{1}\cos\alpha^{\prime}_{t}}\,.
\end{equation}
%
\\
Using the above reflection coefficients we now write the reflected and transmitted electric field expressions for each plane wave by adding s and p components of $E_{ref}^{\prime}$ and $E_{trans}^{\prime}$ respectively as in Section \ref{eq:flat}, 
\begin{equation}
 \vec E^{\prime}_{ref}= \vec E^{\prime s}_{ref}+\vec E^{\prime p}_{ref}\nonumber 
\end{equation}  
\begin{equation}
 \vec E^{\prime}_{trans}= \vec E^{\prime s}_{trans}+\vec E^{\prime p}_{trans} \,.
\end{equation}
\\
 We remind the reader that the new coordinate system $(x^{\prime}-y^{\prime}-z^{\prime})$ is not fixed, rather it changes with the location Q on the spherical surface
which, in turn, depends on the parameters of the plane wave. So we need to
write our final expression in the fixed coordinate system $(x-y-z)$.
 Using the inverse of the rotation matrix $Rot$, 
we finally write the electric field expression in the original coordinate system as 
 
  \begin{equation}
 \vec E_{ref}= Rot^{-1} \cdot \vec E^{\prime}_{ref}
\nonumber 
\end{equation} 
  \begin{equation}
 \vec E_{trans}= Rot^{-1} \cdot \vec E^{\prime}_{trans}\,. \nonumber
 \end{equation}
 Since we are interested only in the perpendicular component, we consider only the y-component of the electric field. For each plane wave we obtain the y-component of $\vec E_{ref}$ as 
\begin{equation}\label{eq:erefsp}
  E_{ref, y}= 
\frac{1}{2} {ik^3\over 8\epsilon \pi^2} \tilde \Pi_{S,r} \left[f^{\prime s}_{r}(1+\cos2\beta)
-f^{\prime p}_{r}\cos\alpha\cos(2\alpha^{\prime}-\alpha)(1-\cos2\beta)
\right]\,.
\end{equation}
Similarly we write the y- component of $\vec E_{trans}$ as 
\begin{equation}
  E_{trans, y}= 
\frac{1}{2} {ik_{1}^3\over 8\epsilon_{1} \pi^2} \tilde \Pi_{S,t} \left[f^{\prime s}_{t}(1+\cos2\beta_{t})
+f^{\prime p}_{t}\cos\alpha_{t}\cos(\alpha-\alpha^{\prime}+\alpha^{\prime}_{t})(1-\cos2\beta_{t})
\right]\,.
\end{equation}
We compute the y-component of the total reflected field by integrating Eq. \ref{eq:erefsp} over $d\Omega= \sin\alpha d\alpha d\beta$. 
It is convenient to define $\lambda = k\sin\alpha$. 
We divide the integral over $\lambda$ into 3 regions: 
(i)  $0\le \lambda < \frac{kR}{R+h}$,
(ii)  $ \frac{kR}{R+h}\le \lambda < \frac{k_1R}{R+h}$,
(iii)  $ \lambda \ge \frac{k_1R}{R+h}$.
 Region (i) gives the dominant contribution. The contribution from other
two regions is found to be negligible.
\begin{figure}[!ht]\begin{floatrow}
\ffigbox[\FBwidth]{\includegraphics[scale=0.3]{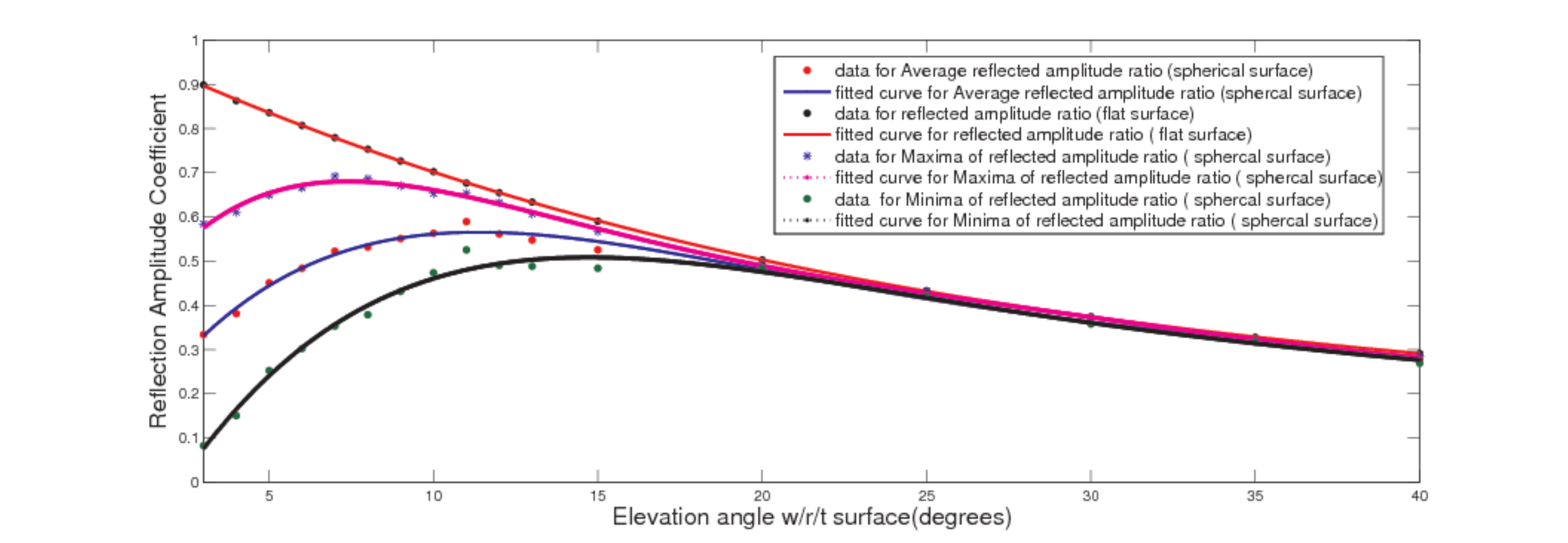}}
{\caption{\it The calculated (n=1.4, f=200 MHz, spherical reflection, no roughness) ratio $\sqrt{r/d}$, with $r$ and $d$ the reflected and direct power respectively, as a function of elevation angle relative to the surface, Since the ratio oscillates rapidly with elevation angle, here we show only the maxima, minima and the average of these oscillations. The data points show the result of a direct numerical calculation and the smooth curves are obtained by 
making a polynomial fit. The result for a flat surface is shown for comparison.}\label{fig:spherical_smooth}} 
\ffigbox[\FBwidth]{\includegraphics[scale=0.48]{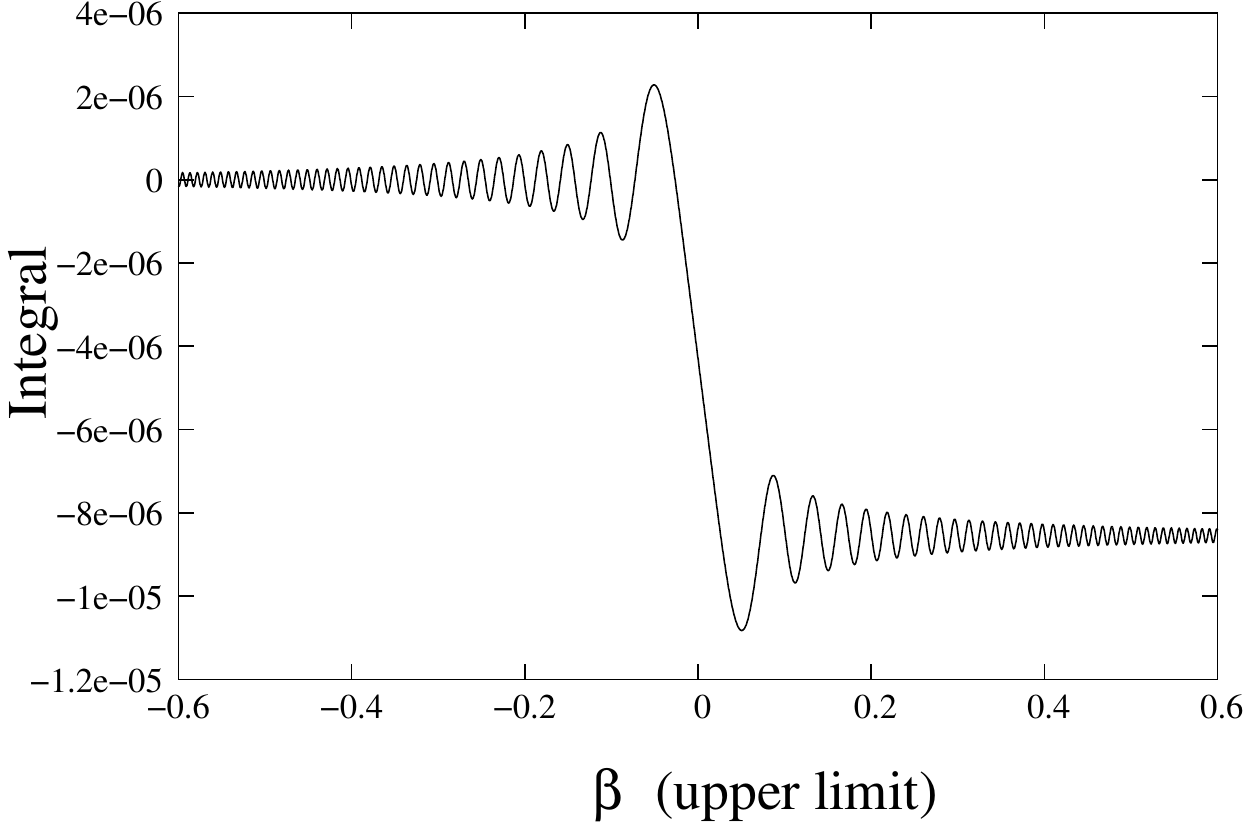}}{\caption{\it Dependence of the integrated real part of the electric field on the integration upper limit on the angle $\beta$. The lower limit has
been set equal to a large negative value. In this case the elevation angle has been set equal to 10 degrees. We have set f = 500 MHz, n = 1.4 and the roughness as given in text.
}\label{fig:integral}} 
\end{floatrow}\end{figure}
The result obtained for the amplitude ratio, choosing a surface index-of-refraction $n=1.4$ and frequency $f=200$ MHz, is shown in Fig. \ref{fig:spherical_smooth}. The result for the flat surface is shown for comparison. This result is relatively insensitive to frequency and shows only a mild increase with the refractive index for small values of elevation angle. Since the ratio oscillates rapidly with elevation angle, we show only the maxima, minima and the average of these oscillations in Fig. \ref{fig:spherical_smooth}. For comparison with data, we should use the mean value. In any case, as discussed below, once roughness corrections are included, the fluctuations disappear.

\subsection{Roughness Correction}
\label{eq:roughness}
The roughness contribution is computed by using the model \citep{romero2015passive}
\begin{equation}
 F(k,\rho,\theta) = \exp\left[{-2k^2\sigma_h(\rho_\perp)^2\cos^2\theta_z}\right]
\end{equation}
where $\rho_\perp^2 = x^2+y^2$ and
\begin{equation}
\sigma_h(L) = \sigma_h(L_0)\left({L\over L_0}\right)^H\,.
\end{equation}
We choose the parameters $L_0=150$ m, $\sigma_h(150 m) = 0.041$ m
and $H=0.65$ which are found to give reasonable agreement with data. We find the power reflectance ratio for a spherical rough surface. In this case we do not observe any oscillations and the power ratio varies smoothly with elevation angle. We find that the contribution is obtained dominantly from a
small region close to the specular point (this is true in the zero roughness case, as well) and hence we can confine the integration to this region. We see this explicitly in Fig. \ref{fig:integral} which shows the integral as a function of the upper limit on the azimuthal angle $\beta$.
The same is found for the case of angle $\alpha$ where the dominant contributions arise from a small region around $\beta\approx 0$. The oscillations seen in Fig. \ref{fig:integral} arise due to the change in path lengths as we integrate over angle $\beta$ across the different fresnel zones.

As discussed below, our numerical result for a spherical rough surface deviates from HiCal2 data for small elevation angle. The deviation from the data can be attributed to our assumption that for each incident plane wave, the corresponding reflected wave is also a plane wave. A more general treatment is under development that does not rely on this assumption. Here we also use an alternate formalism in which the curvature correction is incorporated as a geometric factor.


\subsection{Alternative treatment for calculating power reflectance for a spherical surface}
We can also incorporate the effects due to curvature of Earth by using the divergence factor $D$ \citep{balanis1997antenna} given by, 
\begin{equation}
 D\approx [1+\frac{2s s^\prime}{Rd\tan\psi}]^{-\frac{1}{2}}
\end{equation}
where $\psi$ is the  reflection angle (with respect to the tangent at the point of reflection).\\
$R=$ radius of Earth \\
$d =$ arc length along the surface of Earth between the source and observation point ($OB$) as given in  Fig. \ref{fig:sph1}. \\
$s =$ distance between the specular point and the observation point.\\
$s^\prime =$ distance between the source and the specular point.\\

We compute the flat surface amplitude reflectance including the roughness correction given in \ref{eq:roughness}. This result is then multiplied by the divergence factor $D$. The square of this result gives the final power reflectance. As discussed below, this treatment gives better agreement with HiCal2 data for all elevation angles.

\section{Experimental Results}
To determine
the reflectivity from the interferometric maps formed from ANITA event triggers, we follow three, parallel strategies, and interpret 
the scatter between the three results as a measure of the inherent systematic errors. In each case, we initially 
select event pairs with trigger times 
consistent with the time separation expected for (Reflected,Direct) (designated as
``(R,D)'') pairs, and geometrically
consistent with
the known sky location of HiCal to within 3 degrees in azimuth and also within 3 degrees in elevation, either
above the horizon (direct events) or below the horizon (reflected events). The directional
ANITA interferometric source reconstruction relies on excellent
channel-to-channel timing resolution ($<$100 ps) to find the pixel in the sky interferometric map most consistent with the
measured relative arrival times for received signals. From high Signal-to-Noise Ratio (SNR) data taken while a ground pulser was transmitting from the
Antarctic West Antarctic Ice Station (WAIS), the typical resolution in azimuth $\phi$ and
elevation $\theta$ is determined to be of order $\sigma_\phi\sim0.2^\circ$ and $\sigma_\theta\sim0.4^\circ$.

Once the candidate sample has been chosen by pointing and timing, we
evaluate the reflectivity ${\cal R}$ as follows:
\begin{enumerate}
\item ${\cal R}$ determined from coherently-summed, de-dispersed waveforms:
\subitem For the sample of both R and D events, we form the coherently-summed waveform (i.e., the summed waveform of those channels used to form the interferometric map, after shifting each waveform by the time delay expected for that sky pixel found to give the maximum total cross-correlation),
after deconvolving the system response. The coherently-summed, deconvolved 
waveform is now Fourier transformed into the frequency domain, and the D, or R power calculated in each bin of incidence angle, 
summing over the ``good'' 200--650 MHz system bandwidth for ANITA-4.
\item ${\cal R}$ determined from raw waveforms -- here, we follow the same procedure as used for the HiCal-1 analysis, namely:
\subitem Identify the ANITA-4 antennas pointing to within 45 degrees of the HiCal payload, then calculate the noise-subtracted HPol power in each antenna (summing
the squares of the voltages, and using the first 64 samples
in the captured waveform, prior to the onset of the received signal to measure noise) separately for R vs. D.
\item ${\cal R}$ calculated from the slope of R vs. D:
To ensure that our calculated ratio is insensitive to either trigger threshold biases for low-amplitude reflected events, or saturation effects for high-amplitude direct events, we
plot the square of the peak of the maximum Hilbert transformed voltage for R vs. D, and fit the slope of this graph over the central interval to the form $R={\cal R}D$, constraining
the fit to pass through the point (0,0). Owing to the rotation of the transmitter
payload, even in the case of perfect resolution, the signal strength will vary from event-to-event. 
\end{enumerate}

\subsection{Corrections}
Corrections must be applied to the `raw' values of ${\cal R}$ given by the above prescriptions, as follows.
\subsubsection{Receiver Cant Angle Correction}
The ANITA-4 receiver antennas are canted at 10 degrees below the horizontal to favor reception of upcoming signals
resulting from in-ice neutrino interactions. This results in a calculable correction, as a function of 
incidence angle, for D vs. R events, assuming 
a beam-width $\sigma$=26 degrees for the ANITA Seavey Quad-Ridge receiver horn antennas. 

\subsubsection{Pathlength correction}
There is a straightforward 
correction due to the 1/r diminution of the electric field strength with distance from the source, corresponding to
$1/r^2$ diminution in power, which is different for R vs. D. 
This correction can be as large as 25\% at separation distances smaller than $\sim$150 km.


\subsubsection{Azimuthal correction}
\begin{figure}[htpb]\begin{floatrow}
\ffigbox[\FBwidth]{\includegraphics[width=0.5\textwidth]{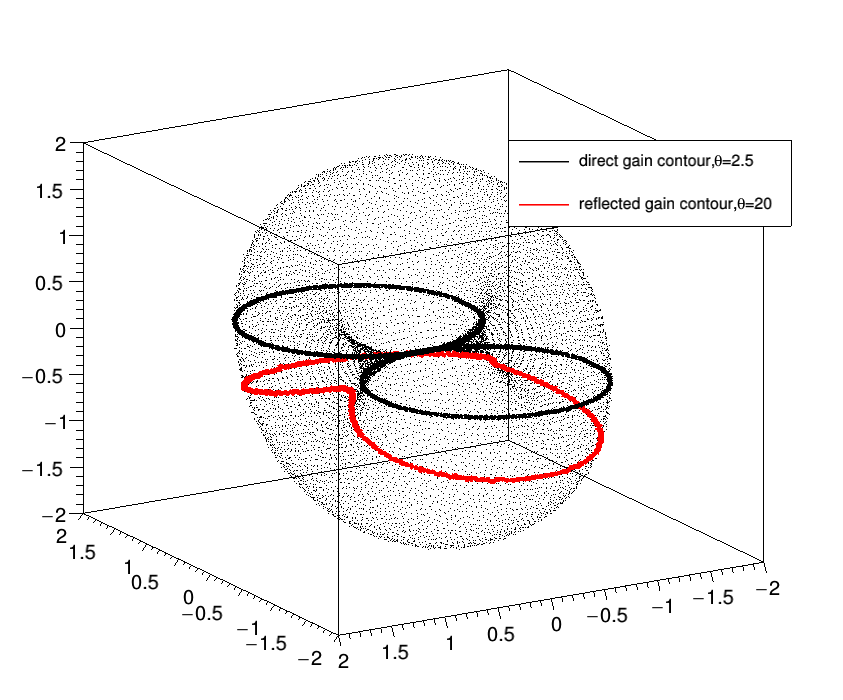}}{\caption{\it Polar gain pattern dependence of dipole on azimuth.}\label{fig:gainpattern}} 
\end{floatrow}\end{figure}

Since the beam pattern of the bicone transmitter roughly follows
$\sin\theta$, with $\theta$ the angle between the signal emission
direction and the transmitter antenna axis,
the gain when antenna boresight
is rotated by a given azimuthal angle relative to ANITA
is different for the D vs. R signals -- in the limit where boresight points perpendicular to
ANITA (i.e., ANITA lies along the antenna axis), e.g., the D gain for HPol signals should be zero, whereas the R HPol gain is non-zero, since the R signal
is emitted at some separation-dependent angle below the horizontal plane (Fig. \ref{fig:gainpattern}).
In such a case, the D signal is (in principle) pure VPol.

\subsection{Cross-checks}
\subsubsection{Check of signal polarization}
We have conducted several cross-checks of our observed signals. The most direct cross-check of true reflected
signals vs. direct signals is the expected $\pi$ radian phase change upon reflection, in the case where the
reflecting surface has a higher index-of-refraction than the medium from which the initial signal is incident --
this is true of both HPol, as well as VPol electromagnetic waves. 
To test this, we compare the correlation coefficient when we
cross-correlate the observed putative reflected signal with the direct signal, vs. an `inverted' (in this case,
by taking the negative of the actual recorded waveform voltages)
reflected signal cross-correlated with the direct signal.
As with HiCal-1, we find that cross-correlation favors the inverted, rather than non-inverted signals.

\subsubsection{Possibility of `ripple' signals}
Owing to imperfect impedance mismatch over the full frequency band of the bicone antenna,
the large, multi-kV piezo-electric signal induced across the antenna feedpoint can result 
in `ringing' that persists considerably longer than the 110 ns timescale of a typical ANITA-4 event capture. Additionally,
the oscillatory relaxation of the piezo can result in after-pulses, separated by several hundred ns.
Since the ANITA-4 buffer depth allows only a maximum of four waveforms stored in memory at a given time,
this raises the possibility of registering
an initial direct event, followed by successive direct `echoes' over the next few microseconds, and thereby initiating
a full system clear and reset. The timescale for the reset ($\sim$10 ms) suppresses the registration of the
reflected signal by ANITA.

\begin{figure}[htpb]\begin{floatrow}
\ffigbox[\FBwidth]{\includegraphics[width=0.5\textwidth]{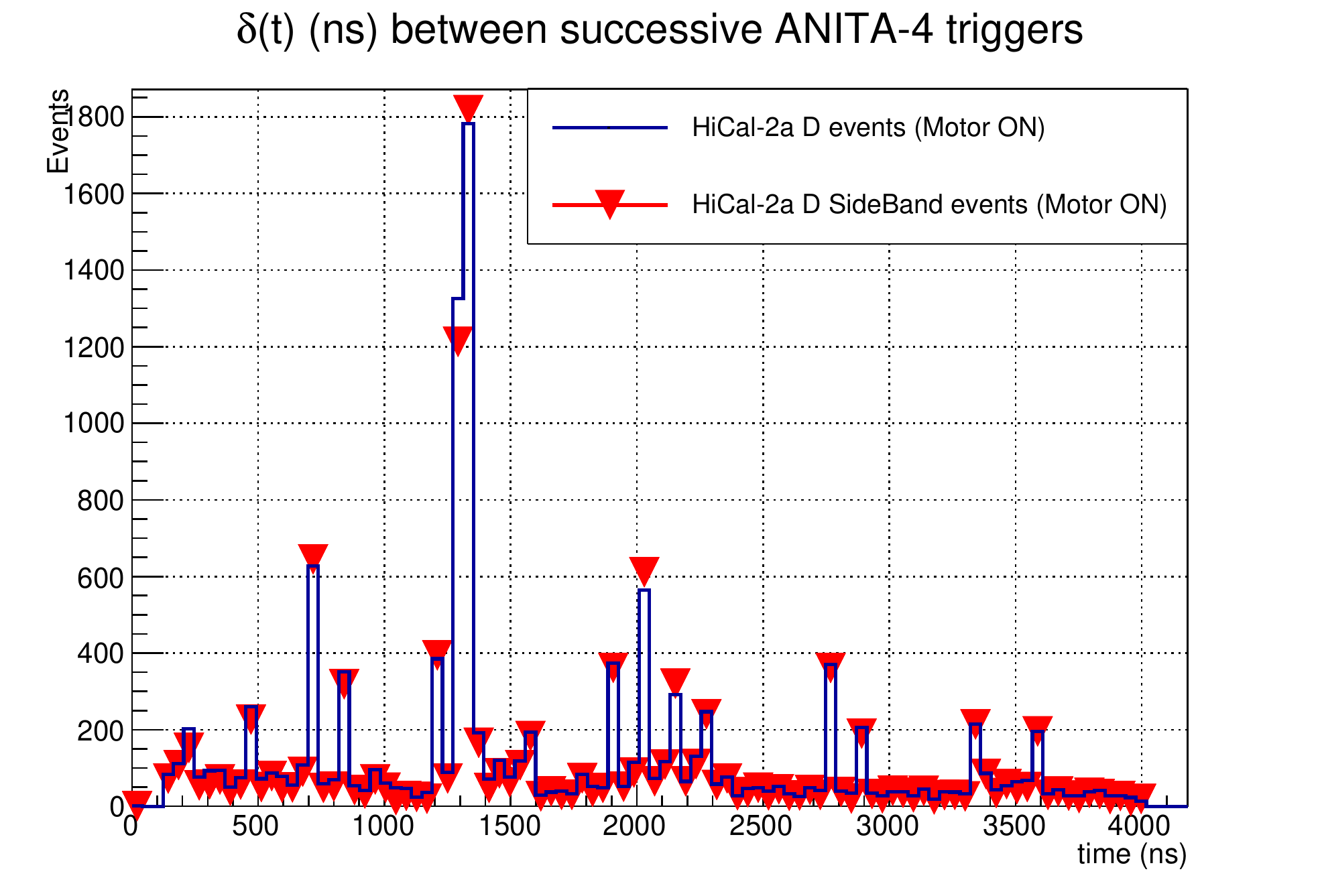}}{\caption{\it Time difference between successive ANITA-4 triggers for HiCal-2a D-events (histogram) vs. estimated HiCal-2a D-background (triangles).}\label{fig:HC2adt}} 
\ffigbox[\FBwidth]{\includegraphics[width=0.5\textwidth]{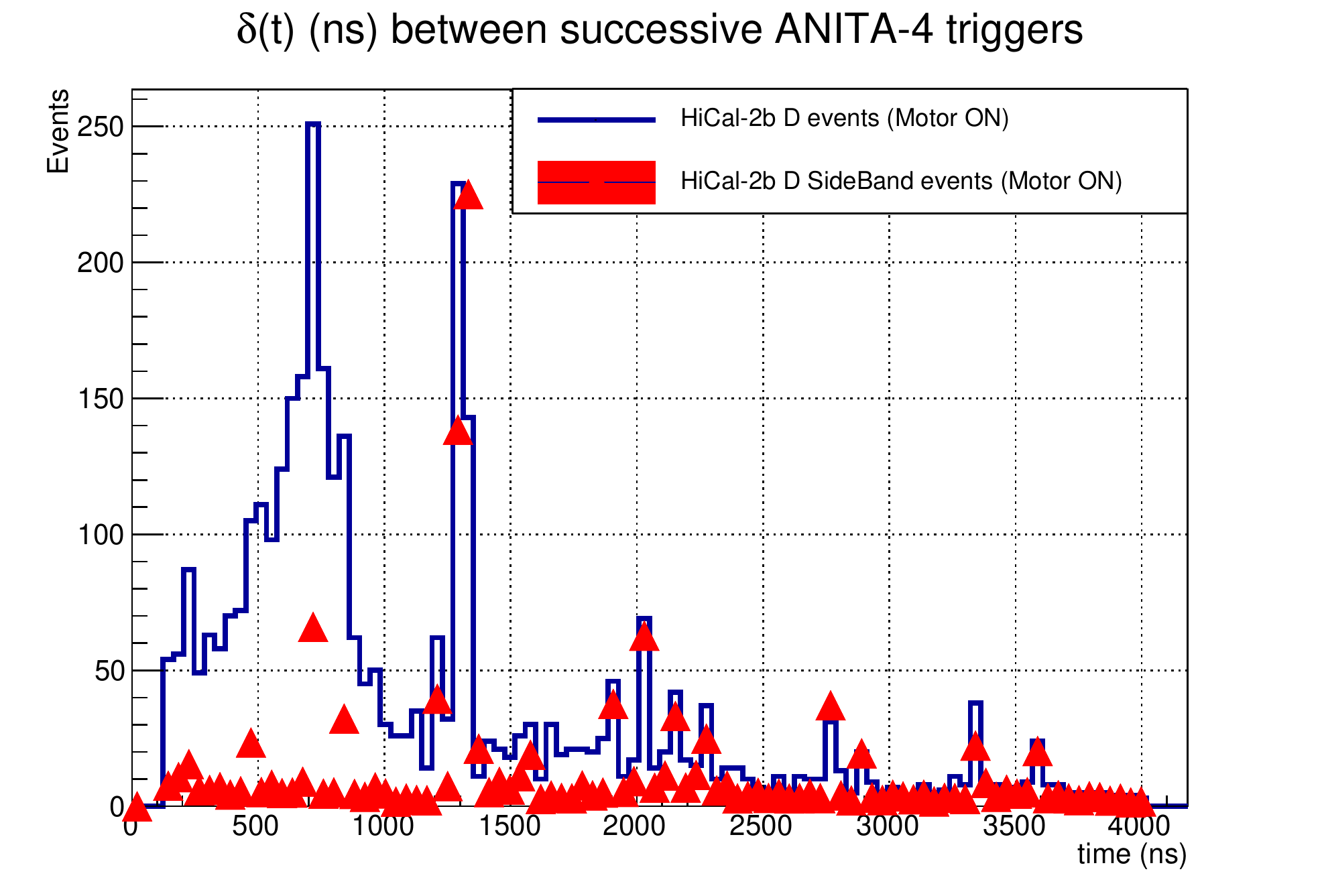}}{\caption{\it Time difference between successive ANITA-4 triggers for HiCal-2b D-events (histogram) vs. estimated HiCal-2b D-background (triangles).}\label{fig:HC2bdt}} 
\end{floatrow}\end{figure}

For HiCal-1b, such an effect was searched for using the sample of 100 (D,R) pairs by considering the angular difference between a putative D event and the previous event trigger, with
no such obvious effects observed. To investigate this for 
HiCal-2b vs. HiCal-2a, we plot the time between successive triggers $\delta(t_{i,j})$ for a) cases where the HiCal piezo was
active, and for which there is a candidate D event identified by pointing, 
vs. b) cases where the HiCal piezo was active, but there is no evident HiCal D event.
As shown in Figures \ref{fig:HC2adt} and \ref{fig:HC2bdt}, restricting consideration to
$\delta(t_{ij})$ values smaller than the minimum possible R-D time difference, we observe
a considerable excess of evident ``echoes'', relative to background, for HiCal-2b compared to HiCal-2a, consistent with
secondary pulses observed from the HiCal-2b piezo pre-flight, and clustering around a period of 600--700 ns. We attribute the bulk of the
observed unpaired D-events to this effect, with the remainder due to cases where the D-event fills the fourth
available buffer, initiating a reset prior to registration of the corresponding R-event. Fortunately, these
events can be readily suppressed in software by requiring that a) the time difference, measured at the ANITA-4 payload, 
between the recorded R event and the
putative D predecessor be consistent with expectation, knowing the elevation of HiCal-2b and ANITA-4 and the separation distance, and b)
explicitly suppressing events pointing directly at HiCal-2b, for which the previous event also points directly at HiCal-2b.

\subsubsection{Check of ANITA-4 pointing resolution}
If we require that the observed time difference between an ANITA-4 trigger and a HiCal-2 event be consistent with
the calculated signal transit time between ANITA and HiCal, we can then measure the angular pointing
resolution of the ANITA-4 gondola relative to the HiCal biconical transmitter source, 
as shown in Figure \ref{fig:dphi}, indicating a resolution (FWHM) better than one degree, 
slightly worse than the resolution obtained from ground pulser data. 
Note that this includes both direct, as well as reflected events, both of which evidently follow
a Gaussian distribution with relatively little indication of non-Gaussian tails.

\subsubsection{Check of transmitter antenna beam pattern}
\begin{figure}[htpb]\begin{floatrow}
\ffigbox[\FBwidth]{\includegraphics[width=0.45\textwidth]{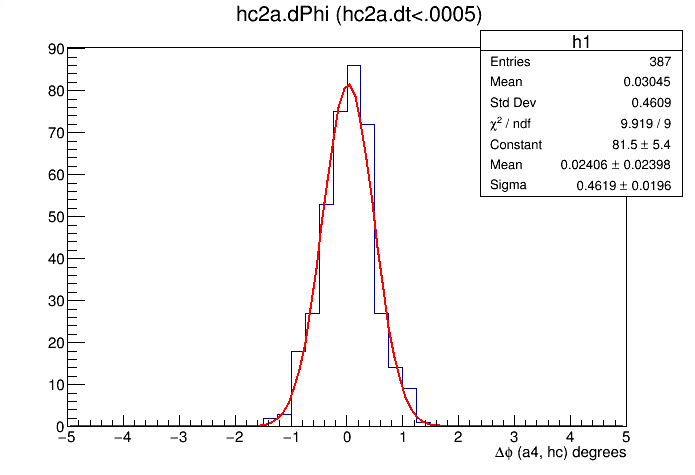}}{\caption{\it Difference between reconstructed source azimuth and true source azimuth for HiCal-2a events, including both direct as well as reflected events. }\label{fig:dphi}} 
\ffigbox[\FBwidth]{\includegraphics[width=0.45\textwidth]{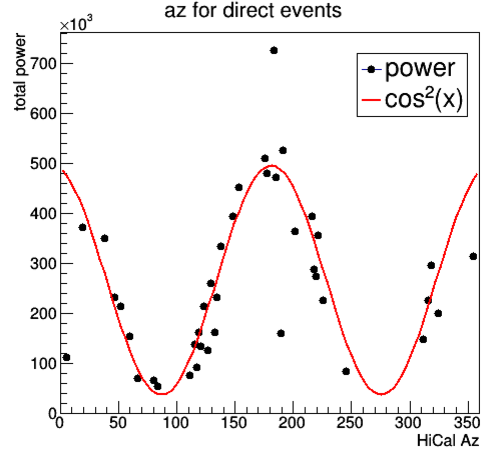}} 
{\caption{\it Recorded HiCal-2 direct signal power, as a function of tabulated HiCal-2 azimuth, overlaid with expected functional variation.}\label{fig:Signal_v_azimuth}} 
\end{floatrow}
\end{figure}

Our bicone transmitter antenna is expected to follow a $\sin\theta$ signal amplitude distribution, measured relative
to the antenna axis. This corresponds to a $\sin^2\theta$ signal power distribution, as derived from
the interferometric map, which itself is computed as the summed product of signal amplitudes. As shown in Figure
\ref{fig:Signal_v_azimuth}, we observe generally adequate match to expectation. We note that the phase of the
overlaid fit has been, in this case, fixed, so there are no free parameters in the fit.


\subsubsection{Trigger Threshold Considerations}The last of our three signal extraction techniques is intended to safeguard against possible trigger threshold effects, since at low signal amplitudes, there may be a possible bias which preferentially selects D/R pairs for which the D power is just above the trigger threshold, but suppresses D/R pairs for which the R power falls just below the trigger thresold. In the previous analysis, this effect was studied using the observed D and R events, and verifying that both distributions were well-separated from the trigger threshold, as defined by the power distribution for thermal triggers. It was additionally tested by verifying that the ratio of R signal to D signal power was linear for all measured pairs.

\begin{figure}[!ht]\begin{floatrow}
\ffigbox[\FBwidth]{\includegraphics[scale=0.45]{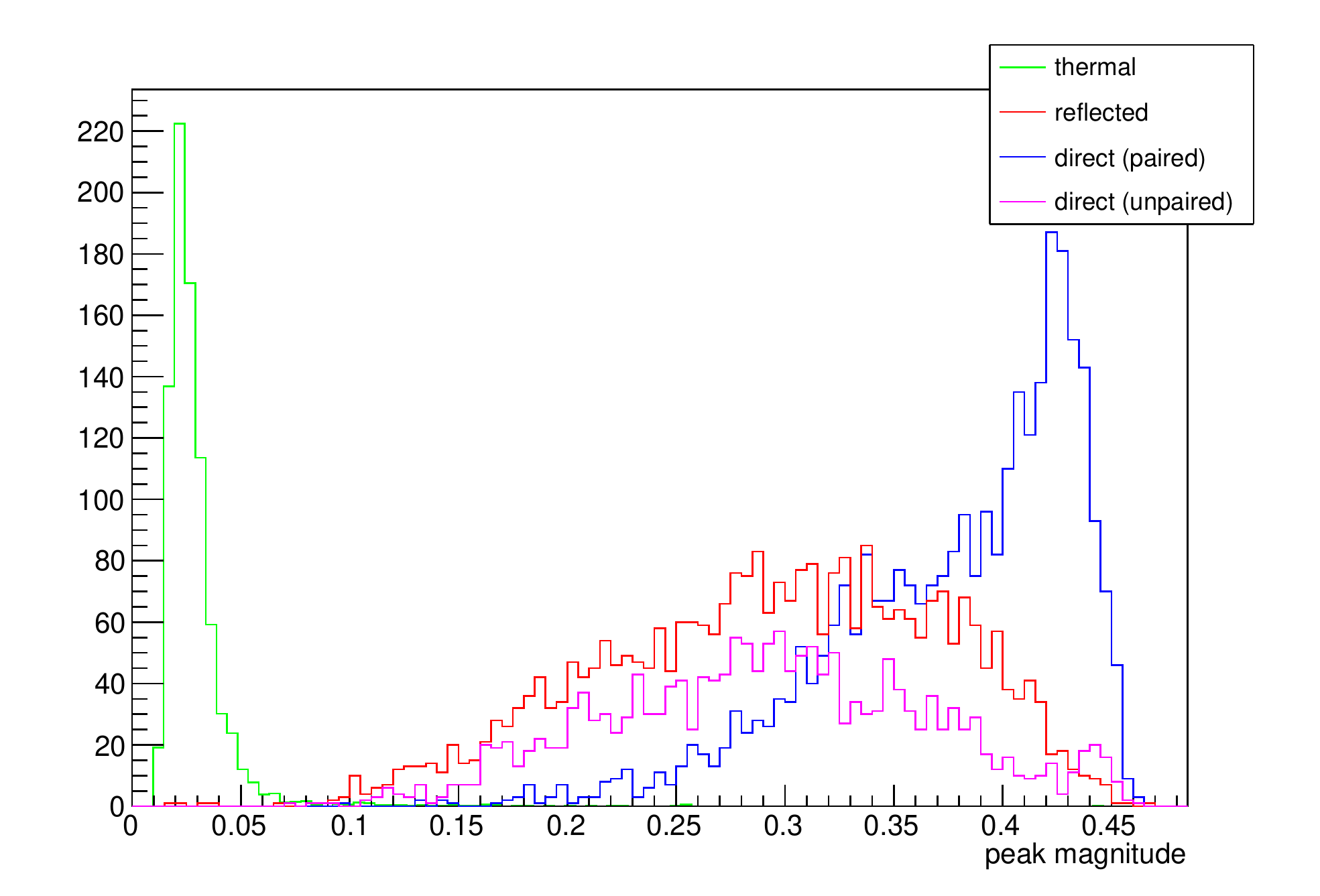}}
{\caption{\it Comparison of signal power for paired D events and paired R events, overlaid with
unpaired D events and also a sample of thermal noise triggers. We observe that the paired R distribution
is well-separated from the thermal noise distribution.}
\label{fig:DRpairpower}}
\ffigbox[\FBwidth]{\includegraphics[width=0.5\textwidth]{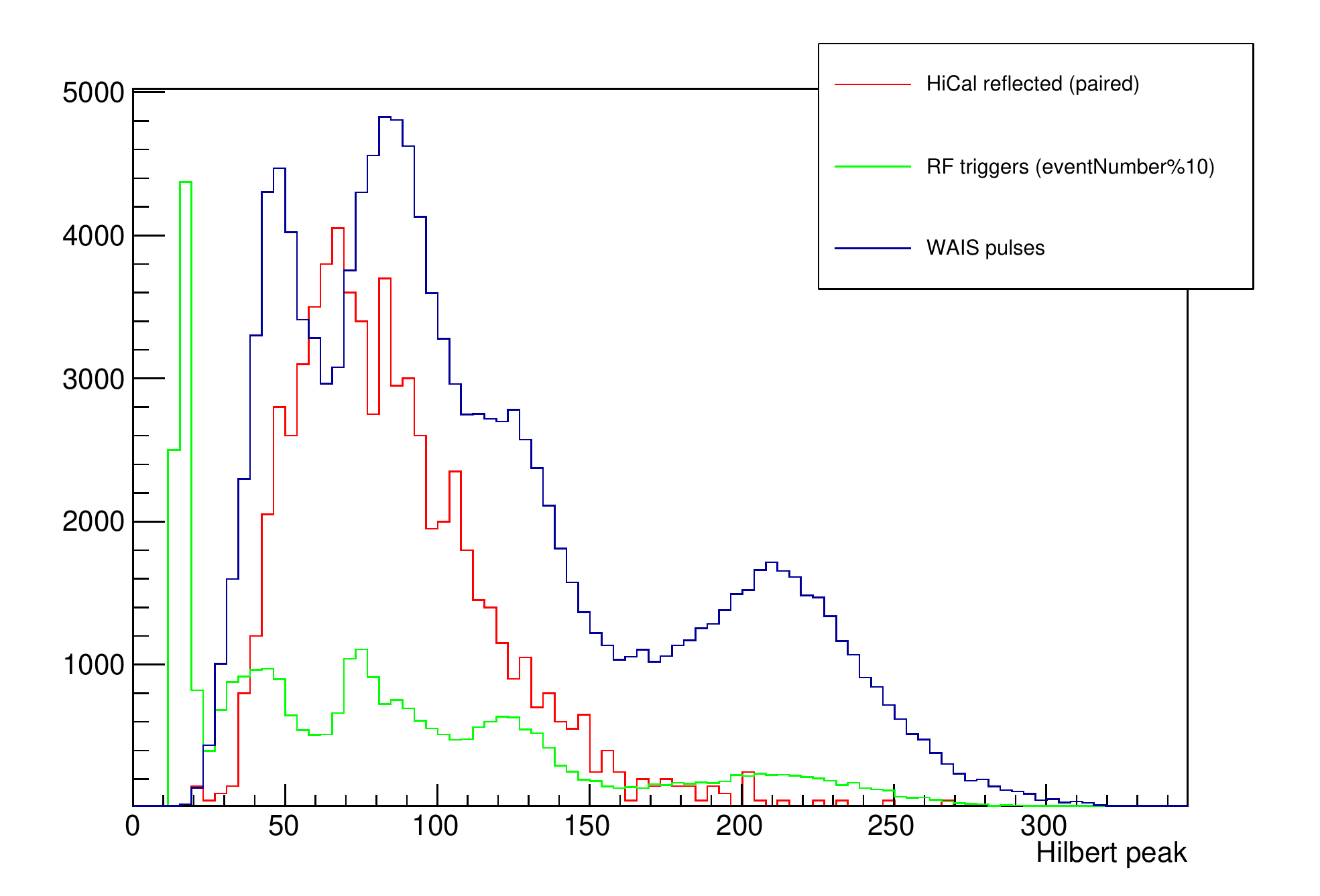}}{\caption{\it Signal strength comparison for reflected events with WAIS calibration pulser sample.}\label{fig:WAIScal}}
\end{floatrow}\end{figure}

With HiCal-2, there are sufficient statistics to study this in greater detail and compare the power distribution for R/D paired events, as well as
D events which are unpaired. For cases where ANITA-4 is off-boresight, or the separation distance between ANITA-4 and HiCal-2 large, 
the D signal will be correspodingly reduced, and the R signal may
be sub trigger-threshold, resulting in an artificially `inflated' measurement.
The paired R signal, however, on which our measurement is based, is found to be well-separated from thermal triggers (Fig. \ref{fig:DRpairpower}). 
As an additional check, we compare the paired R distribution with ground calibration
data taken using a transmitter pulser at the West Antarctic Ice Sheet (WAIS) station (Figure \ref{fig:WAIScal}), again indicating signal strengths well-separated
from threshold.
As a final check, we show the plot of reflected power vs. direct power, for the angular interval showing the greatest discrepancy between measurement and calculation (5--10 degrees incidence with respect to the surface) in Figure \ref{fig:hilbertRD}. We observe saturation at the highest values of Direct power, although we do not observe an obvious deviation from linearity close to the origin. The lack of similar saturation at high values of Reflected power is due, at least in part, to the buffer depth
limitations mentioned previously.

\subsection{Further Probes of Reflectivity}
\begin{figure}[!ht]\begin{floatrow}
\ffigbox[\FBwidth]{\includegraphics[width=0.5\textwidth]{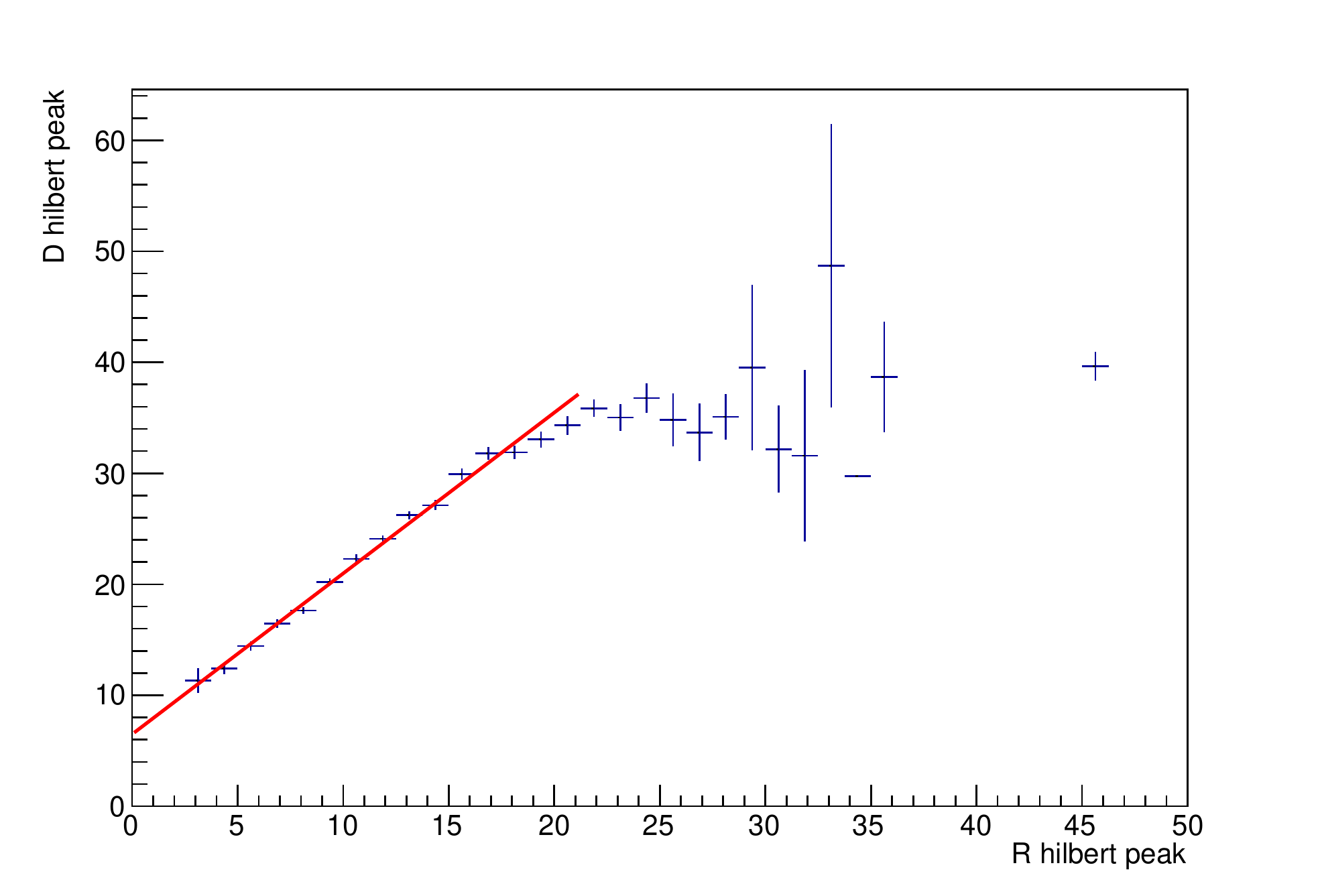}}{\caption{\it Reflected Power vs. Direct Power for events in 5--10 degree angular regime.}\label{fig:hilbertRD}}
\ffigbox[\FBwidth]{\includegraphics[width=0.5\textwidth]{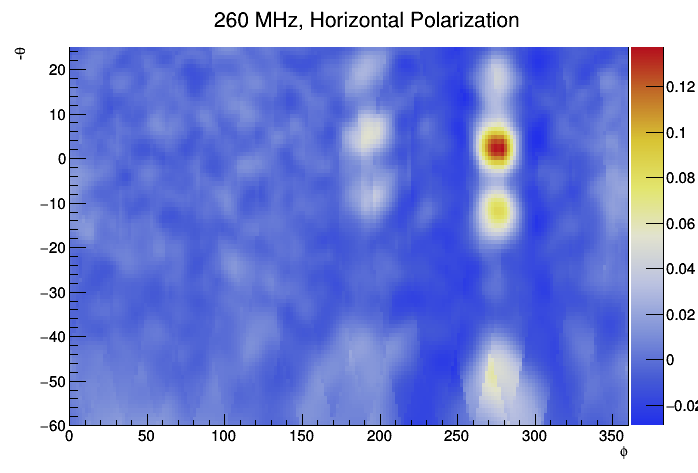}}{\caption{\it ANITA-3 interferometric reconstruction of military satellite broadcasting at 260 MHz (HPol). Direct and surface-reflected signals are observable both above and below the horizontal ($0^o$ in the Figure).}\label{fig:Satellite}}
\end{floatrow}\end{figure}

Thus far, surface reflectivity has been probed using the two HiCal missions and also using continuous solar emissions as the RF source. The ratio of direct signal power in HPol to VPol in the two cases is approximately 10:1 and 1:1, respectively. The former comprises a triggerable, ${\cal O}(10)$ ns signal, while the latter is (obviously) constant and immune to trigger threshold effects. At the time ANITA-3 flew, one of the most pernicious backgrounds was that due to US military satellites, broadcasting at both 260 MHz and 370 MHz, with an SNR comparable to HiCal. Introduction of adaptive frequency filtering in ANITA-4 (``TUnable Frequency Filtering'', or ``TUFF''\citep{allison2017dynamic}) successfully suppressed this background; nevertheless, the narrow band nature of these satellites offers the possiblity of determining the surface reflectivity at a single, fixed frequency value.
As shown in Fig. \ref{fig:Satellite}, 
we can clearly see both the direct and reflected signals due to these satellites; the inferred values of HPol reflection coefficient are: a) 0.52$\pm$0.17 for 260 MHz ($\theta_t\sim 8^\circ$) and b) 0.35$\pm$0.15 for 370 MHz ($\theta_i\sim 6^\circ$). These values are preliminary-only and are presented, at this stage, 
only as a semi-quantitative cross-check of the HiCal-2
results presented herein. 

\subsection{Results Summary}
\begin{figure}[htpb]
\begin{floatrow}
\ffigbox[\FBwidth]{\includegraphics[width=1\textwidth]{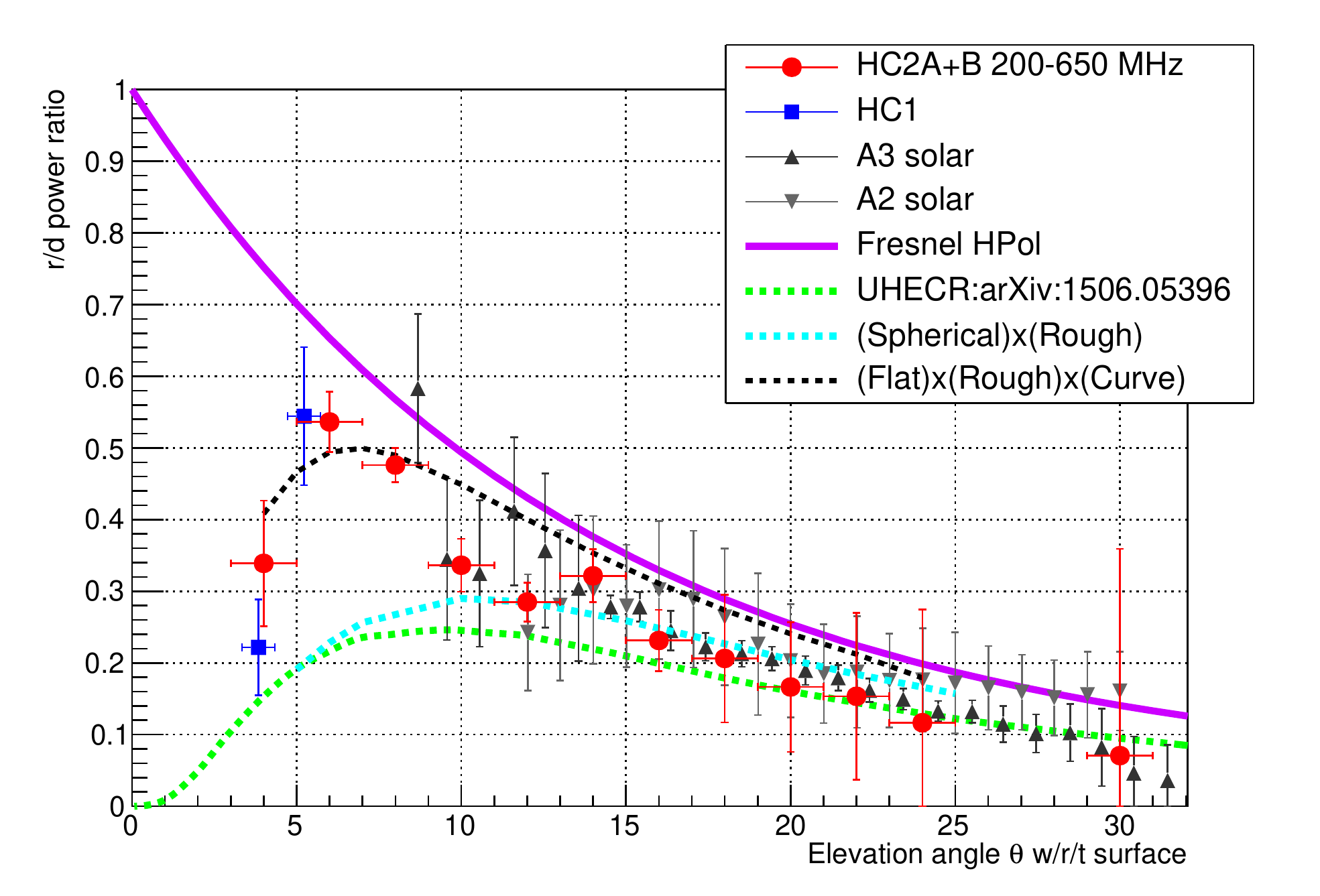}}
{\caption{\it Summary of HiCal results, compared with HiCal-1b results and calculation. As with HiCal-1 and solar measurements, error bars correspond to the widths of R/D distributions and are taken to be indicative of the scale of inherent systematic errors.}\label{fig:Summary}} 
\end{floatrow}\end{figure}

Our reflectivity results are summarized in \ref{fig:Summary}. We note generally good agreement between the
HiCal-2a and HiCal-2b flights and reasonable agreement with the results, at highly oblique incidence angles,
obtained with the HiCal-1b mission. We also note that, the distinct difference in the nature of the emission
(pulsed vs. continuous) notwithstanding, the HiCal-2 results also follow the general trend 
traced by measurements of the 
Solar RF signals (both direct, and reflected), as obtained with both the ANITA-2 and also ANITA-3 experiments.
The black dashed line shown in the
Figure corresponds to our flat-surface calculation including a roughness
correction with the curvature contribution\citep{balanis1997antenna} included with a
multiplicative divergence factor discussed in section 4.4. The roughness parameters used in
this calculation are $L_0 =120$ m, $H =0.65$ , $\sigma_h(L_0) = 0.051$ m, and
the frequency has been set equal to 240 MHz.
The cyan
curve uses the spherical-surface calculation described in section 4.2 including the same roughness
correction. In this calculation, we average over the frequency range 200 to 650 MHz and the roughness
parameters used are $L_0 =150$ m, $H =0.65$ , $\sigma_h(L_0) = 0.041$ m which have been chosen
to provide reasonable agreement with data for elevation angles larger than 10 degrees.
Although in agreement with data over the bulk of the relevant angular regime, at small elevation angles, our own spherical surface calculation, modulo roughness, still underestimates the reflected signal power relative to data.

We note at least one physical difference between the reflector modeled for the purposes of calculation, and the
actual physical reflector -- namely, the reflecting boundary layer is not uniform; HiCal signals penetrating up to 2 wavelengths into the
snow will contribute to the final, observed triggered events. This corresponding 2--3 meter depth of snow also includes seasonal `crusts' with a
local dielectric contrast of order 0.001, which will act as discrete reflecting layers. 
A full Huygens-wavelet inspired simulation of these effects is currently underway.

\section{Outlook and Summary}
Data accumulated with the HiCal-1 and HiCal-2 missions has allowed a fairly comprehensive mapping of the
HPol Antarctic surface reflectivity, over the range of incidence angles relevant to radio-based UHECR measurements.
Of greater relevance to neutrino detection, however, is the vertical polarization surface transmissivity, which
can be inferred as the complement to surface reflectivity. Five obvious goals for a future HiCal-3 mission are
as follows: a) equip the payload with an ADC capable of measuring HPol signal returns at normal incidence from
the surface and provide reflectivity data independent of ANITA-5, b) include solar power provision to extend the lifetime of the measurements, and also offer the possibility of surface reflectivity measurements over sea water, c)
tie the transmitter signal to the GPS second using a triggerable pulser design based on 
a fast DC$\to$DC step-up conversion, d)
inclusion of VPol data collection capabilities, and e) attitude (i.e., polar angle) orientation monitoring. These goals could be met either through two
dedicated missions (as with HiCal-2), or by flying a larger payload-capacity balloon. The timescale of
the ANITA-5 flight (December, 2020) should allow ample time for the HiCal-3 hardware development.

\section{Acknowledgments}
We thank NASA for their generous support of ANITA and HiCal, 
and the National Science Foundation for their
Antarctic operations support. 
specifics required for the HiCal launches relative to ANITA.
 This work was also supported
by the US Dept. of Energy, High Energy Physics Division, 
as well as NRNU, MEPhI and 
 the Megagrant 2013 program of Russia, via agreement 14.А12.31.0006 from 24.06.2013.
We are enormously indebted to the
Columbia  Scientific  Balloon  Facility  for  their excellent support, at all stages of this effort.
Given the specific requirements for the HiCal launch,
particular recognition is due Hugo Franco, Dave Gregory and the entire McMurdo CSBF logistical staff.

\bibliography{Zref}
\bibliographystyle{unsrt}


\end{document}